\documentclass[12pt, a4paper]{article}
\usepackage{amsmath, amssymb}
\usepackage{graphicx}
\usepackage{booktabs}
\usepackage{hyperref}
\usepackage{cleveref}
\usepackage{microtype}
\usepackage{fancyhdr}
\usepackage{amssymb}
\usepackage{tabularx}
\usepackage{booktabs}
\usepackage{ragged2e} 

\usepackage{geometry}
\usepackage{array}
\usepackage{booktabs}
\usepackage{float}
\usepackage{caption}
\usepackage{cite}
\usepackage{amsmath}

\usepackage{graphicx}
\usepackage{enumerate}
\usepackage{array}

\usepackage{multirow}
\usepackage{comment}
\usepackage{mathtools}

\usepackage{amsfonts}
\usepackage{url}
\usepackage{amsthm}
\usepackage{threeparttable}

\usepackage{algorithm}
\usepackage{algpseudocode}
\usepackage{tikz}

\usepackage{booktabs} 
\usepackage{tabularx}

\usepackage{natbib}

\title{Beyond Content: How Author Network Centrality Drives Citation Disparities in Top AI Conferences}

\usepackage{authblk}

\author[1]{Renlong Jie\thanks{Corresponding author. Email: jierenlong@nwpu.edu.cn}}
\author[1]{Longfeng Zhao}
\author[2]{Chen Chu\thanks{Corresponding author. Email: chuchenynufe@hotmail.com}}
\author[3]{Danyang Jia}
\author[3]{Zhen Wang}
\affil[1]{School of Management,
Northwestern Polytechnical University}
\affil[2]{School of Statistics and Mathematics,
Yunnan University of Finance and Economics}
\affil[3]{School of Artificial Intelligence, OPtics and ElectroNics (iOPEN),
Northwestern Polytechnical University}
\date{}

\begin{document}
\maketitle

\begin{abstract}
While scholarly citations are pivotal for assessing academic impact, they often reflect systemic biases beyond research quality. This study examines a critical yet underexplored driver of citation disparities: authors’ structural positions within scientific collaboration networks. Through a large-scale analysis of 17,942 papers from three top-tier machine learning conferences (NeurIPS, ICML, ICLR) published between 2005 and 2024, we quantify the influence of author centrality on citations. Methodologically, we advance the field by employing beta regression to model citation percentiles, which appropriately accounts for the bounded nature of citation data. We also propose a novel centrality metric, Harmonic Closeness with Temporal and Collaboration Count Decay (HCTCD), which incorporates temporal decay and collaboration intensity. Our results robustly demonstrate that long-term centrality exerts a significantly stronger effect on citation percentiles than short-term metrics, with closeness centrality and HCTCD emerging as the most potent predictors. Importantly, team-level centrality aggregation, particularly through exponentially weighted summation, explains citation variance more effectively than conventional rank-based approaches, underscoring the primacy of collective network connectivity over individual prominence. Integrating centrality features into machine learning models yields a 2.4\% to 4.8\% reduction in prediction error (MSE), confirming their value beyond content-based benchmarks. These findings challenge entrenched evaluation paradigms and advocate for network-aware assessment frameworks to mitigate structural inequities in scientific recognition.
\end{abstract}


\maketitle

\section{Introduction} \label{Sec:1}

Citation counts serve as a widely adopted metric for assessing the influence of research papers and play a critical role in evaluating researchers, journals, and conferences. While often interpreted as indicators of scholarly impact, a growing body of literature reveals systematic inequalities and biases embedded in citation practices \citep{kwon2022rise}. These biases manifest at multiple levels, including gender disparities in peer citations \citep{lerman2022gendered}, the under-citation of women-authored papers in fields such as physics \citep{teich2022citation}, and racial as well as geographical underrepresentation in editorial roles \citep{liu2023non}. At a macro level, evidence points to rising cumulative citation inequality alongside increasing collaboration \citep{nielsen2021global}. These patterns collectively demonstrate that structural factors beyond scholarly contribution can significantly shape citation outcomes, which is a concern echoed in critiques of potential biases and misuse inherent in bibliometric indicators \citep{adams2014bibliometrics}. Therefore, it is crucial to identify and disentangle factors beyond research quality that influence citation performance.

Among these, an author’s structural position within scientific collaboration networks (e.g. network centrality) represents a key structural factor. For a given paper, the centrality of some authors may be unrelated to the quality or contribution reflected in its content, yet it can substantially elevate the paper’s citation count. A related mechanism is the prevalence of honorary, guest, and ghost authors across disciplines \citep{flanagin1998prevalence, wislar2011honorary, anstey2014authorship, morreim2023guest}, who typically contribute little intellectually to the work. Since many such authors possess relatively high network centrality, the citation advantage they confer may exacerbate inequities in academic recognition. This pathway itself warrants closer investigation.

Although prior studies have developed models to explain centrality-driven citation effects \citep{yan2009applying, biscaro2014co, guan2017impact}, three major limitations persist. First, their models exhibit limited explanatory power, suggesting omitted variables or misspecification. Second, they fail to quantify the explanatory and predictive gain by centrality metrics given the textual content (e.g., title, abstracts) controlled. Third, the sensitivity of results to author-list aggregation methods (e.g., averaging vs. max centrality per paper) remains unaddressed. These gaps collectively constrain our understanding of how network position inequitably shapes scholarly impact.

To address these limitations, this study conducts a large-scale statistical analysis to examine the effect of authors’ network centrality (measured at the time of publication) on subsequent citation performance. The major contributions of this paper are summarized as follows.
\begin{itemize}
\item We introduce a novel centrality metric and use beta regression to model citation percentiles, offering a precise method to analyze the author centrality–citation relationship, while accounting for the bounded nature of percentile data.
\item By modeling the impact of textual contents with a pretrained transformer-encoder, we isolate the effect of network structure on citation from research contents, providing robust evidence that centrality influence citations beyond paper contents.
\item We show that team-wide centrality aggregation better explains citation variance than first/last-author metrics, highlighting collective team connectivity's key role in driving citations and challenging conventional author-rank evaluation.
\item Through statistical analysis, we find that incorporating a coauthor with an HCTCD centrality 50\% higher than the centrality of the first author will significantly increase the expected citation percentile. 
\item Machine learning models with centrality metrics are developed, showing a 2.4\% to 4. 8\% reduction in prediction error (MSE) compared to content-only models, demonstrating the value of centrality metrics in improving the accuracy of citation prediction.
\end{itemize}

\section{Related Works} \label{Sec:2}

\subsection{Citation Analysis}

Research on citation dynamics reveals significant patterns across disciplines and evolving practices. \citet{radicchi2008universality} demonstrates that while citation distributions vary substantially between fields, they can be rescaled to a universal curve using the relative indicator $c_f = c/c_0$ (where $c_0$ represents the discipline's average citations per article), establishing $c_f$ as an unbiased metric for cross-disciplinary and cross-temporal comparison of citation performance. Complementing this standardization, \citet{varga2019shorter} observes a notable contraction in citation distances between 1950 and 2018, with the average path length decreasing from 5.33 to 3.18 steps, signaling increased cross-field communication and highlighting the critical network-bridging role of high-impact papers, while clustering values for both intra- and inter-field citations show gradual growth. Simultaneously, collaboration analysis by \citet{petersen2015quantifying} identifies ``super ties'' – indispensable, strategically formed partnerships revealed through examination of over 166,000 collaboration records. These intensive collaborations, analyzed from an egocentric perspective across 473 profiles, correlate with a 17\% increase in citations per publication, underscoring their substantial impact on scientific productivity and career advancement. Further refining citation interpretation, \citet{catalini2015incidence} develops a methodology combining bibliometric data and natural language processing to detect ``negative'' citations, finding they frequently target high-quality findings from academically and socially proximate scholars. These advancements collectively reflect the expanding scope of citation analysis, significantly propelled by the availability of large-scale bibliographic databases (e.g., Web of Science, Scopus) and the increasing incorporation of natural language processing and machine learning techniques \citet{iqbal2021decade}.

\citet{yan2009applying} constructs an evolving coauthorship network from 20 years of data across 16 library and information science journals, calculating four centrality measures (closeness, betweenness, degree, and PageRank) and finding they are significantly correlated with citation counts, thus suggesting centrality measures are useful indicators for author impact analysis. \citet{biscaro2014co} examines the effects of co-authorship and bibliographic coupling networks on article citations, finding that authors' degree centrality positively impacts citations and closeness centrality has a positive effect when the giant component is relevant, while betweenness centrality has a negative effect. \citet{guan2017impact} constructs knowledge networks from article keywords and analyzes their impact on citations alongside collaboration networks, revealing that authors' and knowledge elements' network positions significantly affect citation counts. 

\subsection{Citation prediction}

Recent advances in citation prediction leverage diverse methodologies to forecast scholarly impact through network analysis and feature engineering. \citet{yu2012citation} pioneered this domain with a two-phase citation probability model that outperforms conventional link prediction by incorporating author, topic, venue, and temporal features within a topic-discriminative search space. Building on network-based approaches, \citet{sarigol2014predicting} demonstrates that author centrality in coauthorship networks positively correlates with citation success in computer science. Subsequent research identified additional predictive factors: \citet{deng2015papers} revealed an inverse relationship between title length and citation counts, while \citet{bhat2015citation} developed classification models using author interdisciplinarity (quantified via Shannon entropy and Jensen-Shannon divergence), author influence metrics, and lexical features from titles. Recent innovations include \citet{geng2022modeling}'s Dynamic Graph and Node Importance (DGNI) framework that captures evolutionary trends in academic networks to predict citations for new publications, significantly outperforming state-of-the-art models on large-scale datasets. Similarly, \citet{yang2023revisiting}'s CATE-HGN framework advances citation prediction through Cluster-Aware and Text-Enhancing Heterogeneous Graph Neural Networks, demonstrating superior performance in modeling research impact propagation. These methodological developments are systematically cataloged in \citet{xia2023review}'s comprehensive taxonomy, which classifies impact prediction tasks for papers, scholars, venues and institutions into four methodological categories: mathematical statistics, traditional machine learning, deep learning, and graph-based approaches.

\subsection{Centrality Impact}

Beyond the domain of citation analysis, a growing body of research explores how network centrality influences a variety of real-world outcomes across different levels of analysis. Recent meta-analytic work by \citet{nezami2025network}, which integrates findings from 147 studies, offers a comprehensive overview: degree, closeness, betweenness, and eigenvector centrality all exhibit positive effects on firm performance, though their impacts are moderated by temporal, market, and network-specific factors, with eigenvector centrality gaining prominence over time while others diminish.

At the individual level, centrality mediates key performance mechanisms. \citet{ahuja2003individual} find that in virtual R\&D groups, individual centrality fully mediates the effects of functional role and status on publication performance and partially mediates communication effects. Similarly, \citet{he2022executive} show that executives with higher centrality are more prone to financial misreporting, leveraging their influential positions and reduced oversight. Within teams and organizations, centrality operates with nuanced contingency. \citet{kane2008casting} adopt a multimodal network perspective in healthcare teams, revealing that the centrality of information systems significantly predicts care efficiency and quality. In contrast, \citet{sasidharan2012effects} observe that centralized structures inhibit implementation success, though individuals with high in-degree and betweenness centrality report greater task impact and information quality. The evidence from organizational research underscores network centrality as a critical, context-dependent factor in performance. This prompts the question of how such structural advantages operate within scientific collaboration networks to influence citation outcomes, which is a gap our study addresses.

\section{Methods} \label{Sec:3}

\subsection{Collaborative networks} \label{Sec:3.1}

Centrality, as formalized by \citet{freeman1978centrality} and extended to weighted and directed graphs by \citet{stephenson1989rethinking}, provides a theoretically grounded measure of an actor’s structural importance within a collaboration network.
To investigate the impact of author centrality on paper citation, we model the collaborative relationships between authors using complex networks \citep{albert2002statistical, newman2003structure, camarinha2005collaborative}. Specifically, we quantify the co-authorship of an author with other authors through the degree of the network and measure the relative strength of an author in the entire network through centrality metrics. We apply several types of centrality measures, each capturing different aspects of an author's position within the collaborative network \citep{freeman2002centrality, zhang2017degree, page1999pagerank}. In addition, we develop a novel metric called HCTCD and a centrality aggregation method for the author lists. The details of these centrality metrics are given in Section~\ref{Sec:3.1.1} to Section~\ref{Sec:3.1.3}.

\subsubsection{Existing Centrality Metrics} \label{Sec:3.1.1}
\textbf{Degree Centrality}: Measures the number of direct connections (edges) a node has. It reflects the immediate co-authorship relationships of an author. The degree centrality \(C_D(i)\) for author \(i\) is calculated as:
    \begin{equation}
        C_D(i) = \frac{\text{Degree}(i)}{N - 1}
    \end{equation}
    where \(N\) is the total number of nodes (authors) in the network, and \(\text{Degree}(i)\) represents the number of distinct co-authors that author $i$ has collaborated with in all previous publications in the relevant conferences or journals.

\textbf{Closeness Centrality}: Reflects how close a node is to all other nodes in the network. It captures the efficiency with which an author can access all other authors in the network. The closeness centrality \(C_C(i)\) for author \(i\) is calculated as:
    \begin{equation}
        C_C(i) = \frac{N - 1}{\sum_{j \neq i} d(i, j)}
    \end{equation}
    where \(d(i, j)\) is the shortest path length between nodes \(i\) and \(j\) in the collaboration network, $N$ is the total number of nodes in the network or reachable nodes from node $i$. There is also a version of harmonic closeness centrality, which is given by:
        \begin{equation}
        C_{HC}(i) = \frac{\sum_{j \neq i} \frac{1}{d(i, j)}}{N - 1}
    \end{equation}

\textbf{Betweenness Centrality}: Quantifies how often a node lies on the shortest path between other nodes. It measures the extent to which an author acts as a bridge in the network. The betweenness centrality \(C_B(i)\) for author \(i\) is calculated as:
    \begin{equation}
        C_B(i) = \sum_{s \neq i \neq t} \frac{\sigma_{st}(i)}{\sigma_{st}}
    \end{equation}
    where \(\sigma_{st}\) is the total number of shortest paths between nodes \(s\) and \(t\), and \(\sigma_{st}(i)\) is the number of those paths passing through \(i\).

\textbf{Pagerank Centrality}: Originally developed for ranking web pages, Pagerank introduces a damping factor \(d\) to prevent ``importance leakage'' in networks \citep{page1999pagerank}. It has become a fundamental tool for analyzing node importance in both directed and undirected graphs. The Pagerank score \(C_{PR}(i)\) for node \(i\) is computed iteratively as:
    \begin{equation}
        C_{PR}(i) = \frac{1 - d}{N} + d \sum_{j \in N(i)} \frac{C_{PR}(j)}{\text{Degree}(j)}
    \end{equation}
    where \(d \in (0, 1)\) is the damping factor (typically 0.85), \(N\) is the total number of nodes, \(N(i)\) denotes the neighbors of node \(i\), and \(\text{Degree}(j)\) is the degree of node \(j\).

In academic collaboration networks, various centrality metrics offer complementary insights into an author's position and potential influence within the network structure. Degree Centrality measures the number of direct co-authors, reflecting an author's breadth of collaboration but not its depth or indirect influence. Closeness Centrality assesses the average shortest path length to all other authors, indicating overall efficiency in accessing network information and resources, yet it is sensitive to isolated subgroups or fragmented structures \citep{yan2009applying}. Betweenness Centrality highlights an author's role as a bridge between groups, capturing the significance of cross-group collaboration and knowledge flow, though it is computationally demanding and can be dominated by a few key nodes \citep{white1994betweenness, brandes2001faster}. Pagerank Centrality integrates direct and indirect connections, reflecting the process of prestige propagation through iterative computation and a damping factor, making it suitable for measuring long-term accumulated influence in large networks. However, it may underestimate emerging researchers with limited collaborative ties \citep{senanayake2015pagerank, zhang2022analysing}. Collectively, these metrics provide a comprehensive perspective for analyzing an author's network position and potential citation performance.

\subsubsection{Proposed Centrality Metrics} \label{Sec:3.1.2}

Existing centrality metrics fail to fully capture all aspects of collaboration networks necessary for explaining paper citations. To address these limitations, we propose a new centrality measure based on the following assumptions:
\begin{itemize}
\item Collaborations that occurred further in the past have less impact on current citations.
\item Multiple-time collaborations are weighted higher than single-time collaborations, but the increase is not linear. Instead, there is a diminishing marginal effect as the number of collaborations increases.
\item The centrality measure is robust to the presence of isolated nodes or clusters within the network.
\end{itemize}

By taking into consideration of the above assumptions, we develop a parameterized centrality named \textbf{Harmonic Closeness with Temporal and Collaboration Count Decay (HCTCD)}, which is given by Eq.~\eqref{Eq:HCTCD}
    \begin{equation}
        C_{HCTCD}(i) = \frac{\sum_{j \neq i} \frac{1}{d(i, j)}e^{-\alpha(t_p-t_c)+\beta c_{ij}}}{N - 1}
        \label{Eq:HCTCD}
    \end{equation}
First, as the collaborative network involve collaboration across many years, the earlier collaboration or centrality may not contribute as strong as the recent collaboration. Thus, we introduce an exponential decay factor $e^{-\alpha(t_p-t_c)}$, where $\alpha$ is a parameter, $t_p$ is the publication time of paper involving author $i$ for evaluating the citation counts, and $t_c$ is the last collaboration time between author $i$ and author $j$. Thus, the last collaboration time is earlier, the effective distance $d(i, j)e^{\alpha c_{ij}}$ between author $i$ and author $j$ is larger. Also, we consider the collaboration counts between each pair of authors, while the contribution of collaboration counts has a margin effect, we introduce another exponential decay factor $e^{-\beta c_{ij}}$, where $\beta$ is another parameter. The more collaboration times $c_{ij}$, the shorter the effective distance $d(i, j)e^{-\beta c_{ij}}$ between author $i$ and author $j$.

\subsubsection{Aggregation methods for author list} \label{Sec:3.1.3}

As a paper usually involves multiple authors, we need aggregation of centralities of all authors to get the centrality metric for the paper. The aggregation stragety we consider include the first author's centrality, the last author's centrality, the average centrality of all authors, the summed centrality of all authors, the maximum and minimum centralities of all authors, and weighted average/sum centrality of all authors. For weighted centrality, we have the following assumption:

\begin{itemize}
\item The frontier author's centrality get higher weights on the paper citation than the later authors.
\end{itemize}

For the weighted centrality, we apply weights average by exponential decay, which is given by:
\begin{equation}
C_{agg,ave} = \sum_i^n w_i C_i = \sum_i^n \frac{e^{-\tau i}C_i}{\sum_i^n e^{-\tau i}},\quad C_{agg,sum} = \sum_i^n e^{-\tau i}C_i
\label{Eq:Agg}
\end{equation}
where $w_i = \frac{e^{-\tau i}}{\sum_i^n e^{-\tau i}}$ and $\tau$ is a parameter to be estimated, $C_i$ is the centrality of the $i$th author (for any selected centrality metric), $n$ is the number of authors in the author list for the given paper. Here we assume that the first author's centrality is the most important in determining the citation. In this study, we optimize global parameters, including $\alpha$ and $\beta$ for $C_{HCTCD}$, and $\tau$ for $C_{agg}$, with numerical methods to maximize the correlation with the citation percentile among papers published in the same year. The tuned parameters are applied in the centrality metrics for further statistical analysis.
    

\subsection{Beta Regression} \label{Sec:3.2}
To analyze the association of author centrality on paper citation, we use the percentile of citation among papers published in the same year as the dependent variable \(y\). The percentile \(\text{pcite}_i\) for paper \(i\) is calculated as:
\begin{equation}
    \text{pcite}_i = \frac{1}{|S_{t_i}|} \sum_{j \in S_{t_i}} 1\{c_j \leq c_i\}
\end{equation}
where \(c_i\) and \(c_j\) are the citation counts of paper \(i\) and paper \(j\), respectively, and \(S_{t_i}\) is the set of all papers published in year \(t_i\). The citation percentile is uniformly distributed between 0 and 1. The original citation count can be approximately determined given the percentile and the full distribution of the citation count. Exploratory study has shown that the percentile metric is more robust than the Z-score metric (as in \citet{guan2017impact}) in regression analysis.

Given that the value of the percentile is bounded between 0 and 1, we apply a beta regression model to fit the data \citep{ferrari2004beta, liu2018review}. The beta regression model is particularly suitable for modeling continuous rates and proportions. For an instance $(X_i, y_i)$, the corresponding density function is given by:
\begin{equation}
    f(y_i; \alpha_i, \beta_i) = \frac{y_i^{\alpha_i - 1} (1 - y_i)^{\beta_i - 1}}{B(\alpha_i, \beta_i)}
\end{equation}
where \(\alpha_i = \mu_i \phi > 0\) and \(\beta_i = (1 - \mu_i) \phi > 0\) are shape parameters, $\phi$ is the global precision parameter, and \(B(\alpha, \beta)\) is the Beta-function defined as:
\begin{equation}
    B(\alpha, \beta) = \frac{\Gamma(\alpha) \Gamma(\beta)}{\Gamma(\alpha + \beta)}
\end{equation}
where \(\Gamma(\cdot)\) is the Gamma function \citep{sebah2002introduction}. Consider the dataset $(\mathbf{X}, \mathbf{y})=\{(X_1, y_i),..., (X_n, y_n)\}$, the expected value \(E(Y | \alpha, \beta) = \frac{\alpha}{\alpha + \beta} = \mu\). Since the dependent variable is percentile, we use a Probit link function to relate the independent variables $\mathbf{X}$ to the expected value $\mathbf{\mu}$ \citep{aldrich1984linear}:
\begin{equation}
    \mathbf{\mu} = E[\mathbf{y} | \mathbf{X}] = \Phi(\mathbf{\gamma X}),\quad \text{Var}(\mathbf{y}|\mathbf{X}) = \frac{\mu (1-\mu)}{1+\phi}
\end{equation}
where $\mathbf{X}$ is the matrix of independent variables, and \(\gamma\) is the vector of parameters, \(\Phi\) is the cumulative distribution function of the standard normal distribution. This ensures the estimated $E[y|X]$ is bounded between 0 and 1. The parameters are estimated using Maximum log-likelihood estimation with the likelihood function given by Eq.(12).
\begin{align}
\ln L(\alpha,\beta; \mathbf{y})& = \sum_{i=1}^{n} \ln f(y_i|\alpha_i,\beta_i) = \sum_{i=1}^{n} [\ln y_i^{\alpha_i -1} + \ln (1 -y_i)^{\beta_i -1} - \ln B(\alpha_i,\beta_i)]  \\
\alpha &= \phi\mathbf{\mu}(\mathbf{\gamma}, \mathbf{X}),\quad\beta = \phi(1-\mathbf{\mu}(\mathbf{\gamma}, \mathbf{X})) \nonumber
\label{Eq:LL}
\end{align}
where $\mathbf{\gamma}$ and $\phi$ are the model parameters to be optimized. We re-parameterize the target as: 
\begin{equation}
\mathbf{\gamma}^{*}, \phi^*=\text{argmax}_{\mathbf{\gamma}, \phi}\ln L(\mathbf{\gamma}, \phi; \mathbf{y}, \mathbf{X})
\end{equation}
The details of the estimation process is given in \citep{ferrari2004beta}. This approach allows us to effectively model the relationship between author centrality and paper citation while accounting for the bounded nature of the dependent variable. 

\subsection{Variables and notations}

The variables and corresponding notations are summarized in Table~\ref{tab:vars}, where ``X'' represents the type of centralities discussed in Section~\ref{Sec:3.1}.
\begin{table}[htbp]
  \centering{
  \caption{Variables and explanations}
    \begin{threeparttable}
    \begin{tabular}{p{3.0cm}p{12.0cm}}
    \toprule
    \textbf{Variable} & \multicolumn{1}{c}{\textbf{Explanation}} \\
    \midrule
    Pcite & The percentile of current citation among the papers published in the same year. \\
    YearToNow & The time between publication date and now in years. \\
    N.Author & The number of authors of the paper. \\
    Len.Abs. & The length of abstract in characters. \\
    Len.Title. & The length of title in characters. \\
    Model & The output of a pre-trained model. \\
    Degree & The variable of degree centrality. \\
    Closeness & The variable of closeness centrality. \\
    Betweenness & The variable of betweenness centrality. \\
    Cpagerank & The variable of pagerank centrality. \\
    HCTCD & The variable of proposed HCTCD centrality. \\
    X.1st & The X-centrality value of the first author. \\
    X.1st.d & Difference of 2-year and 8-year X-centrality values of the first author. \\
    X.Sum. & The sum of X-centrality values of all authors of the paper. \\
    X.Sum.d. & Difference of 2-year and 8-year sum of X-centrality values of all authors of the paper. \\
    X.Last & The X-centrality value of the last author. \\
    X.Last.d. & The difference of 2-year and 8-year X-centrality values of the last author. \\
    X.Ave. & The average of X-centrality values of all authors. \\
    X.Ave.d. & The difference of 2-year and 8-year average X-centrality values of all authors. \\
    X.Max & The maximum centrality values among all authors of the paper. \\
    X.Min & The minimum centrality values among all authors of the paper. \\
    X.W.Ave. & The weighted average of X-centrality values of all authors by Eq.~\eqref{Eq:Agg}. \\
    X.W.Sum. & The weighted summation of X-centrality values of all authors by Eq.~\eqref{Eq:Agg}. \\
    X-n & X-centrality calculated in a n-year window before the publication year of the paper. \\
    \bottomrule
    \end{tabular}
    \begin{tablenotes}
      \item \textbf{Notes:} X is one among Degree/Closeness/Betweenness/Cpagerank/HCTCD.
    \end{tablenotes}
  \end{threeparttable}
  \label{tab:vars}}%
\end{table}%

\section{Experiments and Analysis} \label{Sec:4}

\subsection{Setting Up} \label{Sec:4.1}

\subsubsection{Datasets} \label{Sec:4.1.1}
We collect the meta information of AI top conference paper in Aminer \citep{tang2016aminer, wan2019aminer}, including the title, abstract, author ids, citation, venue, publication year of each paper. We focus on the top-3 machine learning conferences including NeurIPS, ICLR and ICML with 17,942 published papers from 2005 to 2024 that are available in Aminer, including 8,222 papers in NeurIPS, 4,776 papers in ICML and 4,944 papers in ICLR.

\subsubsection{Experimental design}  \label{Sec:4.1.2}

To explore the role of author centrality in citation outcomes, we devise a statistical analysis targeting several key research questions:
\begin{itemize}
\item Do the centrality metrics of the first or last author have a stronger influence on citations than other authors on a paper?
\item How do different centrality metrics (e.g., degree, closeness, betweenness, PageRank, HCTCD) perform in explaining citation variance?
\item How do different methods for aggregating centrality across a paper's author list (e.g., sum, average, max, weighted sum) compare in their ability to predict citations?
\item Given the first author, what is the impact of including a co-author whose centrality is significantly higher (e.g., 50\% higher) than that of the first author on a paper's expected citation count?
\end{itemize}
Methodologically, we dynamically compute centrality metrics for time windows of 1, 2, 4, 8, and 16 years before each paper’s publication year, and calculate temporal derivatives to capture recent centrality changes. Centrality values are aggregated via summation, averaging, and positional extremes at the team level and compared with rank-based metrics. Beta regression models are used to analyze the link between centrality features and citation percentiles while controlling for venue, title/abstract length, and publication year. We also train machine learning models with and without centrality features to assess predictive gains.

For variable specification, the dependent variable is citation percentile (pcite). Independent variables include dynamic window-based centrality, temporal derivatives, and aggregated centrality values. Control variables are venue (NeurIPS/ICLR/ICML), publication year, abstract/title length (Len.Abs, Len.Title), author count (N.Author), and outputs from a transformer-based content model.

In the experiment, author centrality is computed based on co-authorship networks constructed from publications prior to the focal paper’s publication year. Meanwhile,  citations are measured as percentiles among papers published in the same year, accumulated after publication. This design feature mitigates concerns regarding reverse causality that citations could directly affect retrospectively computed centrality, since future citations cannot influence historical network position.

\begin{figure*}[tb]
\begin{center}
 \includegraphics[width=1.0\linewidth]{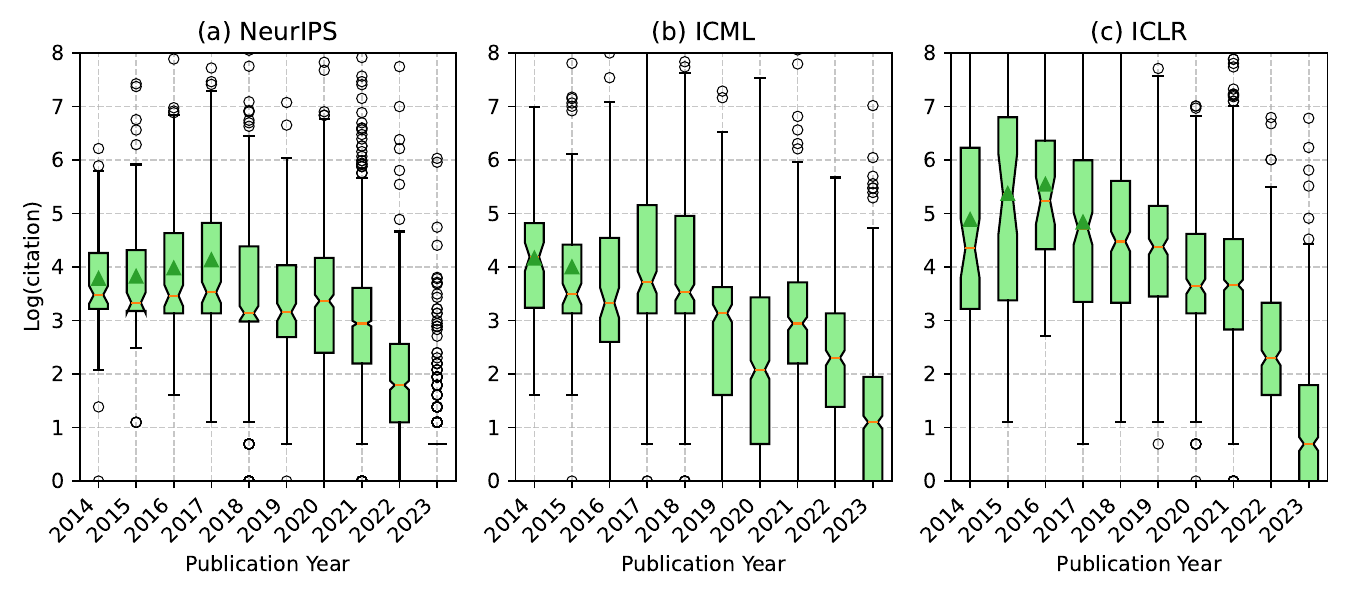}
 \caption{Boxplots of paper citations across venues and years. } \label{Fig:boxplot_citation}
\end{center}
\end{figure*}
\subsection{Exploratory and Visual Analysis} \label{Sec:4.2}

\subsubsection{Visual Analysis} \label{Sec:4.2.1}

We provide the box-plots of logarithms of current citation of papers published in different venues and different years in Figure~\ref{Fig:boxplot_citation}. In general, citation is positively associated with the years after publication. However, the pattern is not linear. The papers published in 2017 get the highest total citation in NeurIPS and ICML, and the papers published in 2015 get the highest citation in ICLR. This indicates breakthrough of novel directions and architectures in artificial intelligent around that time. As the total citation count is strongly but not linearly associated with the publication year index, the selection of year-wise citation percentile as the dependent variable is supported. Meanwhile, we can learn from Figure~\ref{Fig:boxplot_citation} that given the year index, the citation count of different venues differs. For example, the medium citation counts of ICLR papers from 2014 to 2021 are generally higher than that of NeurIPS and ICML papers. This indicate that the venue should be utilized as an independent variable.

\begin{figure*}[t]
\begin{center}
 \includegraphics[width=1.0\linewidth]{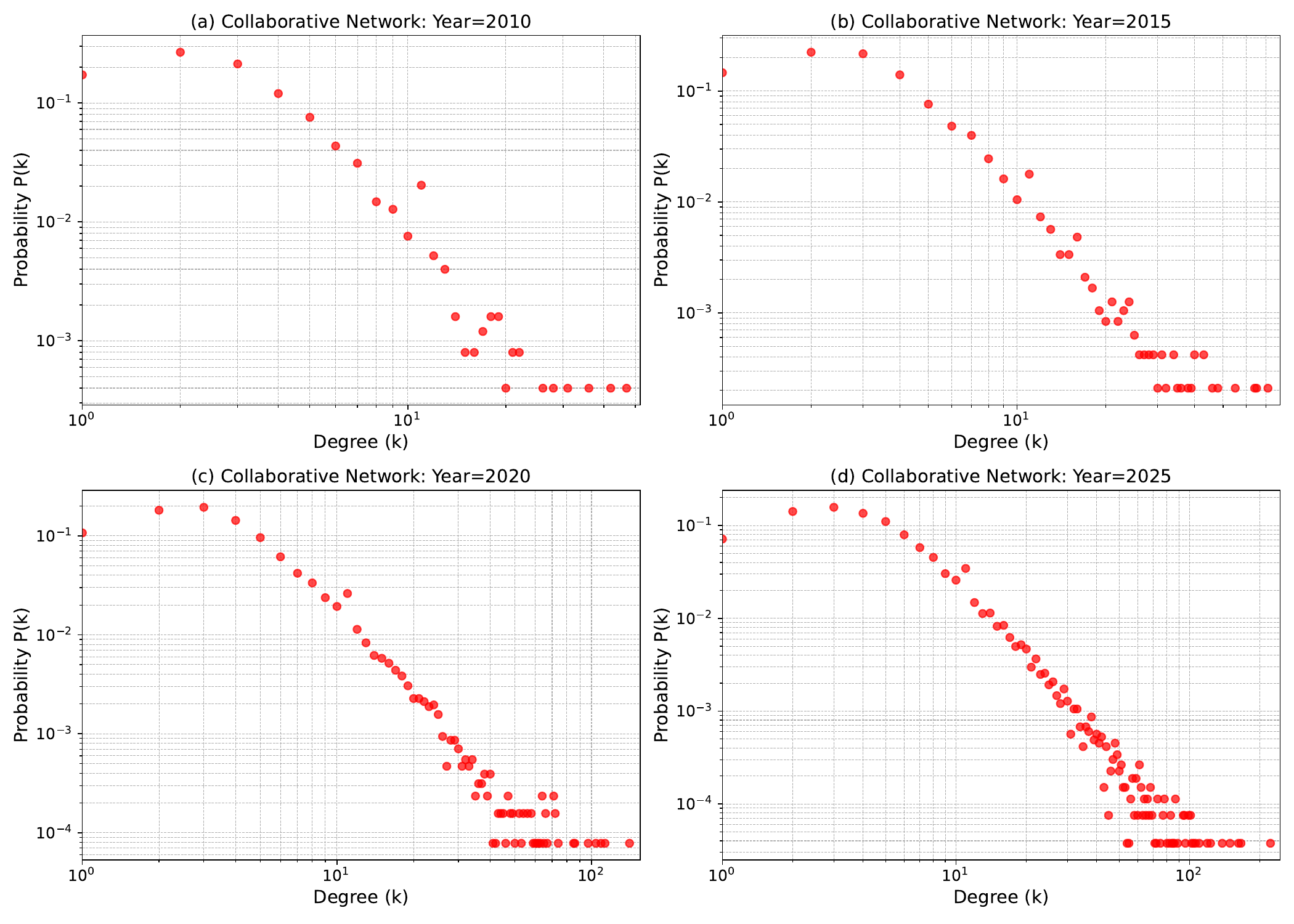}
 \caption{Degree distributions of author networks from 2005 to 2010, 2015, 2020 and 2025. } \label{Fig:degree_dist}
\end{center}
\end{figure*}

We also show the degree distributions of collaborative networks of all these conferences in Figure~\ref{Fig:degree_dist}. We consider the collaborative networks of four different time ranges (2005-2010, 2005-2015, 2005-2020, and 2005-2025) in four subplots and find that they show all linear patterns in log-log coordinates, indicating power-law distributions.  

\subsubsection{Parameter searching for centrality metrics} \label{Sec:4.2.2}

\begin{figure*}[t]
\begin{center}
 \includegraphics[width=1.0\linewidth]{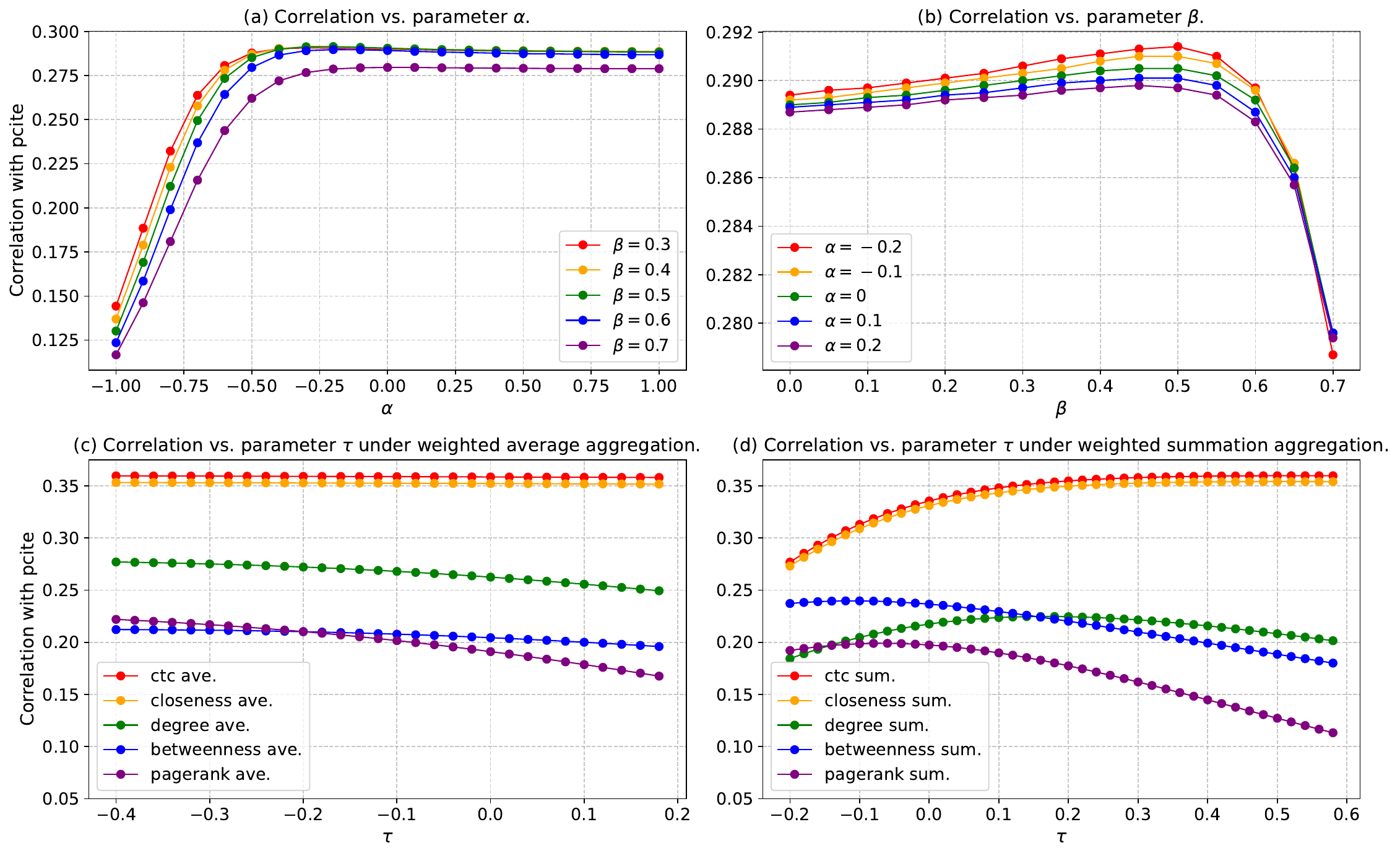}
 \caption{Correlation with percentile of citations given different parameter values. We build these curves using the subset of data from 2015 to 2016.} \label{Fig:parameter}
\end{center}
\end{figure*}

To optimize parameters for centrality metrics, we conducted grid searches on a stratified subset of 2015–2016 publications, maximizing Pearson correlation with citation percentiles (pcite). For PageRank centrality, scanning the damping factor $d \in (0,1)$ in 0.025 increments revealed an optimum at $d=0.975$
(correlation: 0.145), indicating near-complete retention of network influence during random walks (Figure~\ref{Fig:parameter}(a)). For the proposed HCTCD metric, a joint 2D grid search over $\alpha, \beta \in [0,1]^2$ yielded $\alpha=-0.2$ and $\beta=0.45$. The negative $\alpha$ suggesting stronger contributions from older collaborations, potentially reflecting enduring impacts of established authors, while positive $\beta$ confirms diminishing returns for repeated collaborations (Figure~\ref{Fig:parameter}(b)).

Similarly, the author-list aggregation weight $T$ was optimized to $\tau=0.3$ for weighted summation (Eq. 7), implying the second author’s centrality receives approximately 74\% weight ($e^{-0.3}$) of the first author’s, aligning with lead-author prioritization conventions (Figure~\ref{Fig:parameter}(c)). Sensitivity analysis in Figure 3 shows: (a) PageRank’s sharp correlation peak at $d=0.975$, (b) HCTCD’s high-correlation region $\alpha\in [-0.3,0]$, $\beta\in [0.4,0.6]$), and (c) weighted aggregation’s correlation plateau beyond $T=0.3$. These parameters, validated on the full dataset, were fixed for subsequent analyses.



\subsubsection{Temporal Dynamics of Centrality Metrics} \label{Sec:4.2.3}
Analysis of centrality metrics across varying time windows (1–16 years) reveals a consistent pattern wherein long-term network positioning exerts substantially greater influence on citation outcomes than short-term connectivity. As demonstrated in Table~\ref{tab:acorr_time}, all centrality metrics exhibit enhanced correlations with citation percentiles as the temporal window expands, with closeness centrality (both standard and weighted variants) achieving the most pronounced gains: its correlation coefficient rises from 0.313 (1-year window) to 0.389 (16-year window), reflecting a 24.3\% increase. Similarly, degree centrality plateaus at 0.284 for intermediate windows (2–8 years) before declining slightly at 16 years (0.275), while betweenness centrality peaks at 0.222 for a 2-year window before diminishing to 0.175 at 16 years. Notably, the proposed Harmonic Closeness with Temporal and Collaboration Count Decay (HCTCD) metric (HCTCD.W.Sum.) consistently outperforms conventional measures across all windows, culminating in a maximal correlation of 0.397 at 16 years. This temporal progression underscores the compounding effect of sustained scholarly integration, where enduring network embeddedness rather than transient collaborations serves as the primary catalyst for citation visibility. The attenuation of predictive power for short-term metrics (e.g., 1-year closeness centrality correlation decays by 19.5\% relative to 16-year) further corroborates that citation advantage accrues disproportionately to authors with established, persistent network presence.
\begin{table}[t]
  \centering{\small
  \caption{Correlation of different centralities with Pcite on different time spans. All correlation values are w.r.t. Pcite.}
    \begin{tabular}{cccccc}
    \toprule
          & \textbf{1} & \textbf{2} & \textbf{4} & \textbf{8} & \textbf{16} \\
    \midrule
    1.Degree & 0.272  & 0.284  & 0.284  & 0.284  & 0.275  \\
    2.HCTCD   & 0.322  & 0.375  & 0.376  & 0.380  & 0.382  \\
    3.Cpagerank & 0.216  & 0.234  & 0.240  & 0.240  & 0.232  \\
    4.Closeness & 0.313  & 0.367  & 0.373  & 0.378  & 0.389  \\
    5.Betweenness & 0.195  & 0.222  & 0.216  & 0.195  & 0.175  \\
    6.HCTCD.W.Sum. & 0.323  & 0.378  & 0.385  & 0.394  & 0.397  \\
    7.HCTCD.W.Ave. & 0.319  & 0.374  & 0.377  & 0.382  & 0.385  \\
    \bottomrule
    \end{tabular}%
  \label{tab:acorr_time}}%
\end{table}%

\subsubsection{Comparative Centrality Performance} \label{Sec:4.2.4}
\begin{table}[b]
  \centering{\small
  \caption{Correlation of different centralities}
    \begin{tabular}{cccccccc}
    \toprule
          & \textbf{1} & \textbf{2} & \textbf{3} & \textbf{4} & \textbf{5} & \textbf{6} & \textbf{7} \\
    \midrule
    1.Pcite & 1.000  &       &       &       &       &       &  \\
    2.HCTCD-8 & 0.380  & 1.000  &       &       &       &       &  \\
    3.Degree & 0.284  & 0.730  & 1.000  &       &       &       &  \\
    4.Betweenness & 0.195  & 0.569  & 0.841  & 1.000  &       &       &  \\
    5.Closeness & 0.378  & 0.962  & 0.633  & 0.502  & 1.000  &       &  \\
    6.Cpagerank & 0.240  & 0.644  & 0.961  & 0.846  & 0.556  & 1.000  &  \\
    7.N.Author & 0.263  & 0.368  & 0.401  & 0.237  & 0.337  & 0.309  & 1.000  \\
    \bottomrule
    \end{tabular}%
  \label{tab:corr_cent_agg}}%
\end{table}%
The relative efficacy of centrality metrics in explaining citation variance demonstrates significant divergence, with HCTCD centrality emerging as the most effective predictor. Table ~\ref{tab:corr_cent_agg} highlights its superior bi-variate correlation with citation percentiles (0.378), exceeding degree centrality (0.284), PageRank (0.240), and betweenness centrality (0.195). The dominance of closeness centrality, which quantifies global network accessibility, implies that citation advantage accrues primarily to authors positioned to efficiently disseminate knowledge across the collaborative ecosystem, rather than those with high local connectivity (degree centrality) or brokerage roles (betweenness centrality). Furthermore, the high multicollinearity between degree and PageRank centrality ($r=0.961$) suggests redundancy in these local-influence measures when degree centrality is applied, whereas the moderate correlation between closeness and degree centrality ($r=0.633$) indicates complementary informational value. These findings collectively suggests to treat global network efficiency (closeness) as the paramount structural determinant of citation success.

\subsubsection{Aggregation Method Efficacy} \label{Sec:4.2.5}

Team-level aggregation of centrality metrics substantially outperforms rank-based approaches in explaining citation variance, challenging conventional evaluation paradigms that prioritize first or last authors. As quantified in Table ~\ref{tab:corr_cent}, average closeness centrality across all authors (HCTCD.ave) achieves a correlation of 0.381 with citation percentiles, marginally exceeding positional metrics such as first-author centrality (0.379) and last-author centrality (0.370). Crucially, weighted summation (HCTCD.wsum), which exponentially discounts contributions by author order, yields the strongest overall correlation (0.394), surpassing both unweighted summation ($\Delta r=0.024$) and maximum/minimum centrality aggregation (0.380 and 0.381, respectively). This superiority persists in regression analyses (see Table ~\ref{tab:aggregation}), where summed centrality demonstrates the highest statistical significance ($\hat{\gamma}=0.4823, p<0.001$) and model fit ($R^2=0.6092$). The consistent outperformance of collective aggregation methods, particularly weighted summation, validates that citation advantage stems from holistic team connectivity rather than individual prominence. This evidence necessitates a paradigm shift from author-rank-centric evaluation toward network-informed assessment of collective scholarly integration.

\begin{table}[htbp]
  \centering{\footnotesize
  \caption{Correlation of different centrality aggregation methods.}
  \resizebox{\linewidth}{!}{
    \begin{tabular}{crrrrrrrrrrr}
    \toprule
          & \multicolumn{1}{c}{\textbf{1}} & \multicolumn{1}{c}{\textbf{2}} & \multicolumn{1}{c}{\textbf{3}} & \multicolumn{1}{c}{\textbf{4}} & \multicolumn{1}{c}{\textbf{5}} & \multicolumn{1}{c}{\textbf{6}} & \multicolumn{1}{c}{\textbf{7}} & \multicolumn{1}{c}{\textbf{8}} & \multicolumn{1}{c}{\textbf{9}} & \multicolumn{1}{c}{\textbf{10}} & \multicolumn{1}{c}{\textbf{11}} \\
    \midrule
    1.Pcite & 1.000  &       &       &       &       &       &       &       &       &       &  \\
    2.HCTCD.Sum   & 0.370  & 1.000  &       &       &       &       &       &       &       &       &  \\
    3.HCTCD.Ave.  & 0.381  & 0.838  & 1.000  &       &       &       &       &       &       &       &  \\
    4.HCTCD.1st   & 0.379  & 0.838  & 0.991  & 1.000  &       &       &       &       &       &       &  \\
    5.HCTCD.Last  & 0.370  & 0.829  & 0.980  & 0.963  & 1.000  &       &       &       &       &       &  \\
    6.HCTCD.Max   & 0.380  & 0.858  & 0.984  & 0.969  & 0.977  & 1.000  &       &       &       &       &  \\
    7.HCTCD.Min   & 0.381  & 0.836  & 0.996  & 0.992  & 0.971  & 0.971  & 1.000  &       &       &       &  \\
    8.HCTCD.Std   & 0.287  & 0.675  & 0.751  & 0.711  & 0.785  & 0.848  & 0.700  & 1.000  &       &       &  \\
    9.HCTCD.W.Sum.    & 0.394  & 0.928  & 0.971  & 0.968  & 0.951  & 0.967  & 0.968  & 0.740  & 1.000  &       &  \\
    10.HCTCD.W.Ave. & 0.382  & 0.841  & 0.999  & 0.989  & 0.984  & 0.987  & 0.995  & 0.761  & 0.971  & 1.000  &  \\
    11.N.Author & 0.263  & 0.693  & 0.339  & 0.350  & 0.339  & 0.368  & 0.347  & 0.271  & 0.493  & 0.343  & 1.000  \\
    \bottomrule
    \end{tabular}}%
  \label{tab:corr_cent}}%
\end{table}%

\subsubsection{Methodological Implications} \label{Sec:4.2.6}
The high multicollinearity between temporal variants (Table 1: $r > 0.91$ for degree measures) indicates that only one of them (probably the one with the highest correlation with pcite) is needed in the subsequent regression models. The superior performance of closeness centrality underscores the importance of global network positioning over local connectivity, informing the design of our novel HCTCD metric. The aggregation results advocate for team-level centrality analyses rather than traditional author-rank approaches, aligning with emerging trends in collective impact assessment.

\subsection{Regression Analysis}

\subsubsection{Setting up}

To investigate the association between different centrality-related factors and the citation percentile, we build several regression models with the following cases: 

\textbf{Case 1}: The independent variables include different centrality metrics of the authors of a given paper, venue, year, the total number of authors, the title length, and the abstract length. 

\textbf{Case 2}: The independent variables include those in the first case, as well as the outputs of a trained transformer-based model without considering the centrality information. The detail of this model is given in Section~\ref{Sec:4.3.2}, while the textual contents of title and abstract are trained to predict the citation percentile.

\textbf{Case 3}: We apply different aggregation methods to merge the centralities from the author list. Meanwhile, we introduce the difference between 2-year and 8-year centrality of the authors to take account of the increasing or decreasing trends of author centralities.

We focus on the second case as we want to control the effect of the research area and textual content. We use the data samples from 2015 to 2020 for building the regression models, since we want to focus on papers after the revolution of modern deep learning methods, and we believe the citation count within a short period of time is not reflecting the final value and recognition of a paper. In terms of the centrality, we calculated it with a larger window size of 8 years before the date of the sampled papers.

\subsubsection{Model-based baseline} \label{Sec:4.3.2}

We construct a transformer-based model to predict citation percentiles, and utilize its predicted values as a content-control factor in the subsequent regression analyses. Our approach builds upon a pre-trained bert-base-uncased transformer encoder with 12 layers, 12 attention heads, and a hidden size of 768, which we fine-tune on a curated text dataset. The dataset comprises all accepted papers from NeurIPS, ICLR, and ICML, using the title and abstract as input features and the same-year citation percentile as the target variable. The data are split into training, validation, and test sets in an 8:1:1 ratio. An Adam optimizer with a learning rate of $0.00002$ is implemented for model training with an effective batch size of 64.
We evaluate four experimental settings: (a) Use all of venue name, year, title and abstract as the input information. (b) Remove one of the four parts mentioned in (a) and use the rest three as the input information. 
The performance under each configuration, measured on the test set, is summarized in Table~\ref{tab:transformer}. Reported results represent the median values across five independent trials with different random seeds.
In general, the transformer-based citation percentile prediction model demonstrates significant performance variations across metadata configurations. The full model (Setting 1), incorporating venue, year, title, and abstract, achieves optimal performance (Corr = 0.6963, MSE = 0.06844). Exclusion of abstract metadata (Setting 2) marginally reduces correlation ($\Delta$ Corr = -0.0155), while omitting venue information (Setting 4) induces the most severe degradation (Corr = 0.5516, MSE = 0.07163). These results underscore the critical role of venue prestige and textual content in citation prediction. The robustness of title and abstract features (Settings 2-3) aligns with prior findings on content-driven impact assessment, though venue metadata remains indispensable for cross-conference generalization.


\begin{table}[thb]
  \centering{\small
  \caption{The correlation and MSE between the predicted outputs and the citation quartile on test set in different settings.}
    \begin{tabular}{cccccccc}
    \toprule
    Setting	& Year & Venue & Title & Abstract & Corr & MSE & MAE\\
    \midrule
    Setting 1 & \checkmark     & \checkmark     & \checkmark     & \checkmark     & 0.6963 & 0.06844 & 0.2240 \\
    Setting 2 & \checkmark     & \checkmark     & \checkmark     &       & 0.6808 & 0.06925 & 0.2263 \\
    Setting 3 & \checkmark     &       & \checkmark     & \checkmark     & 0.6256 & 0.07075 & 0.2282 \\
    Setting 4 &       & \checkmark     & \checkmark     & \checkmark     & 0.5516 & 0.07163 & 0.2299 \\
    \bottomrule
    \end{tabular}%
  \label{tab:transformer}}%
\end{table}%


\subsection{Beta regression vs OLS}

Given the bounded nature of the dependent variable that the citation percentile (pcite) of papers in the same year is within the $[0,1]$ interval, this study employs a Beta regression framework for modeling \citep{ferrari2004beta}. We apply python package \textit{statsmodels.othermod.betareg} to perform beta regression. To validate the superiority of this method over traditional Ordinary Least Squares (OLS), we constructed two sets of control models: the first introduces the first author's closeness centrality (Closeness.1st), and the second uses the maximum closeness centrality among the author list (Closeness.max). We controlled for covariates such as publication delay (YearToNow), conference venue (ICLR/ICML), abstract length (Len.Abs), title length (Len.Title), and number of authors (N.Author). We apply mean squared error (MSE) to compare the predictive power of both the models.
As shown in Table 6, Beta regression demonstrates superior fitting performance in both model sets: in the first author centrality model, the Beta regression $MSE = 0.05831$ is significantly lower than OLS's 0.05842; in the maximum centrality model, the Beta regression $MSE = 0.05828$ also surpasses OLS's 0.05840. Although the increase in $R^2$ is limited, Beta regression consistently maintains a performance advantage across all settings, indicating its better ability to capture the distributional characteristics of bounded data.
\begin{table}[thb]
  \centering{\footnotesize
  \caption{Comparison of beta regression and linear regression}
    \begin{tabular}{ccccc}
    \toprule
    \multirow{2}[2]{*}{Model} & \multicolumn{2}{c}{Linear Regression} & \multicolumn{2}{c}{Beta Regression} \\
          & 1     & 2     & 1     & 2 \\
    \midrule
    Const &     0.4309*** &     0.4323*** &    -0.2590* &    -0.2518* \\
          &     (0.026) &     (0.026) &     (0.109) &     (0.109) \\
    YearToNow &    -0.0049* &    -0.0049* &    -0.0244* &    -0.0246* \\
          &     (0.002) &     (0.002) &     (0.010) &     (0.010) \\
    ICLR  &     0.1104*** &     0.1102*** &     0.4496*** &     0.4491*** \\
          &     (0.010)&     (0.010 &     (0.039) &     (0.039) \\
    ICML  &    -0.0407*** &    -0.0404*** &    -0.1630*** &    -0.1622*** \\
          &     (0.009) &     (0.009) &     (0.036) &     (0.036) \\
    Len.Abs. &  4.01e-05*** &  4.00e-05*** &     0.0002*** &     0.0002*** \\
          &  (9.77e-06) &  (9.76e-06) &  (4.73e-05) &  (4.73e-05) \\
    Len,Title &    -0.0015*** &    -0.0015*** &    -0.0059*** &    -0.0059*** \\
          &     (0.000) &     (0.000) &     (0.001) &     (0.001) \\
    N.Author &     0.0198*** &     0.0196*** &     0.0742*** &     0.0732*** \\
          &     (0.002) &     (0.002) &     (0.009) &     (0.009) \\
    Closeness.1st &     1.3904*** &       &     5.6081*** &  \\
          &     (0.076) &       &     (0.312) &  \\
    Closeness.max &       &     1.2173*** &       &     4.8975*** \\
          &       &     (0.066) &       &     (0.272) \\
    Precision &       &       &     1.0887*** &     1.0888*** \\
          &       &       &     (0.019) &     (0.019) \\
    \midrule
     MSE & 0.05842 & 0.05840 & \textbf{0.05831} & \textbf{0.05828} \\
    \bottomrule
    \end{tabular}%
  \label{tab:beta_lin}}%
\end{table}%

Theoretical rationale and empirical results jointly support the applicability of Beta regression: First, its Probit link function maps linear predictors to the $(0,1)$ interval, addressing the inherent flaws of OLS in predicting boundary values (see Table ~\ref{tab:beta_lin} for coefficient scale differences: Beta coefficient for Closeness.1st is 5.61 vs. OLS coefficient 1.39). Second, the precision parameter ($\phi \approx 1.09, p < 0.001$), significantly different from zero, confirms the suitability of the Beta distribution for modeling the variability structure of citation percentiles. Finally, although OLS also detects the significance of centrality variables ($p < 0.001$), Beta regression provides a more robust modeling foundation for subsequent analysis through its strict distributional assumptions. Therefore, we establish Beta regression as the core methodological framework for investigating the relationship between network centrality and citation performance.

\subsubsection{Hybrid regression model}

We apply beta regression discussed in Section~\ref{Sec:3.2}. The regression model with selected variables is given by: $Y_i \sim f(y|\alpha, \beta), \, E(Y|\alpha, \beta)=\mu_i,\, \mu_i = \Phi(u_i)$ with
\begin{equation}
\begin{split}
u_i &= \gamma X_i = \gamma_0 + \gamma_1 \text{YearToNow}_i + \gamma_2 \text{ICLR}_i + \gamma_3 \text{ICML}_i + \gamma_4 N.Author \\
& + \gamma_5 \text{Len.Abs}_i + \gamma_6 \text{Len.Title}_i
 + \gamma_7 \text{Model}_i + \gamma_8 X_i + \gamma_9 \text{X.d.}_i + \epsilon_i 
 \nonumber
\end{split}
\end{equation}
where $\gamma = (\gamma_0,...,\gamma_9)$ are the parameters of independent variables. Here we introduce the variable $X.d_i=X_{i2}-X_{i8}$, which captures the change in centrality between the 2-year and 8-year windows, reflecting the recent increase in centrality. Our results demonstrate that centrality metrics provide significant additional explanatory power. 
\begin{table}[t]
  \centering{\footnotesize
  \caption{Summary table of Beta regression on different model settings.}
  \begin{threeparttable}
    \begin{tabular}{p{2.0cm}p{1.7cm}p{1.7cm}p{1.7cm}p{1.7cm}p{1.7cm}p{1.7cm}}
    \toprule
    \textbf{Models} & \textbf{1} & \textbf{2} & \textbf{3} & \textbf{4} & \textbf{5}  & \textbf{6} \\
    \midrule
    \multirow{2}[1]{*}{Const} &   -15.91*** &   -15.55*** &   -15.60*** &   -15.54*** &   -15.74*** &   -15.64*** \\
          &     (0.273) &     (0.277) &     (0.276) &     (0.279) &     (0.275) &     (0.276) \\
\multirow{2}[0]{*}{YearToNow} &    -0.0185* &    -0.0197* &    -0.0341*** &    -0.0229** &    -0.0258** &    -0.0374*** \\
          &     (0.008) &     (0.008) &     (0.009) &     (0.008) &     (0.009) &     (0.009) \\
\multirow{2}[0]{*}{ICLR} &     0.0396 &     0.0279 &     0.0284 &     0.0354 &     0.0387 &     0.0347 \\
          &     (0.033) &     (0.033) &     (0.033) &     (0.033) &     (0.033) &     (0.033) \\
\multirow{2}[0]{*}{ICML} &    -0.0094 &    -0.0070 &    -0.0159 &    -0.0034 &    -0.0140 &    -0.0162 \\
          &     (0.030) &     (0.030) &     (0.030) &     (0.030) &     (0.030) &     (0.030) \\
\multirow{2}[0]{*}{N.Author} &     0.0507*** &     0.0360*** &     0.0318*** &     0.0385*** &     0.0437*** &     0.0378*** \\
          &     (0.007) &     (0.007) &     (0.008) &     (0.007) &     (0.007) &     (0.007) \\
\multirow{2}[0]{*}{Len.Abs.} &  9.88e-5** &  9.49e-5** &  9.52e-5** &  9.57e-5** &   9.88e-5** &  9.43e-5** \\
          &  (3.59e-5) &  (3.57e-5) &  (3.53e-5) &  (3.59e-5) &  (3.58e-5) &  (3.54e-5) \\
\multirow{2}[0]{*}{Len.Title} &    -0.0014* &    -0.0011 &    -0.0012* &    -0.0011 &    -0.0014* &    -0.0012* \\
          &     (0.001) &     (0.001) &     (0.001) &     (0.001) &     (0.001) &     (0.001) \\
\multirow{2}[0]{*}{Model} &    30.46*** &    29.67*** &    30.05*** &    29.66*** &    30.23*** &    30.13*** \\
          &     (0.496) &     (0.508) &     (0.498) &     (0.510) &     (0.497) &     (0.498) \\
\multirow{2}[0]{*}{HCTCD} &       &     1.305*** &       &       &       &  \\
          &       &     (0.197) &       &       &       &  \\
\multirow{2}[0]{*}{Degree} &       &       &    38.35*** &       &       &  \\
          &       &       &     (5.974) &       &       &  \\
\multirow{2}[0]{*}{Closeness} &       &       &       &     1.44*** &       &  \\
          &       &       &       &     (0.229) &       &  \\
\multirow{2}[0]{*}{Betweenness} &       &       &       &       &     8.17*** &  \\
          &       &       &       &       &     (1.792) &  \\
\multirow{2}[0]{*}{Cpagerank} &       &       &       &       &       &   207.8*** \\
          &       &       &       &       &       &    (36.94) \\
\multirow{2}[1]{*}{Precision} &     1.73*** &     1.74*** &     1.74*** &     1.74*** &     1.73*** &     1.73*** \\
          &     (0.020) &     (0.020) &     (0.020) &     (0.020) &     (0.020) &     (0.020) \\
    \midrule
      Log-Likelihood &   1950.6 &   1972.3 &   1971.7 &   1970.1 &   1961.4 &   1966.8 \\
      AIC &   -3883. &   -3925. &   -3923. &   -3920. &   -3903. &   -3914. \\
      BIC &   -3826. &   -3861. &   -3860. &   -3856. &   -3839. &   -3850. \\
    R-squared & 0.6045 & 0.6083 & 0.6081 & 0.6077 & 0.6062 & 0.6074 \\
    \bottomrule
    \end{tabular}%
            \begin{tablenotes}
      \item \textbf{Notes:} ***, **, and * indicate 0.001, 0.01, and 0.05 levels of significance, respectively.
    \end{tablenotes}
  \end{threeparttable}
  \label{tab:centralities}}%
\end{table}%

\begin{table}[htbp]
  \centering{\footnotesize
  \caption{Summary table of Beta regression on different centrality aggregation settings.}
  \begin{threeparttable}
    \begin{tabular}{p{2.0cm}p{1.7cm}p{1.5cm}p{1.5cm}p{1.5cm}p{1.5cm}p{1.5cm}p{1.5cm}}
    \toprule
    \textbf{Models} & \textbf{1} & \textbf{2} & \textbf{3} & \textbf{4} & \textbf{5} & \textbf{6} & \textbf{7} \\
    \midrule
    \multirow{2}[1]{*}{Const} &   -15.55*** &   -15.49*** &   -15.50*** &   -15.57*** &   -15.50*** &   -15.51*** &   -15.53*** \\
          &     (0.278) &     (0.278) &     (0.277) &     (0.278) &     (0.278) &     (0.277) &     (0.277) \\
\multirow{2}[0]{*}{YearToNow} &    -0.0197* &    -0.0140 &    -0.0141 &    -0.0154 &    -0.0138 &    -0.0138 &    -0.0146 \\
          &     (0.008) &     (0.008) &     (0.008) &     (0.008) &     (0.008) &     (0.008) &     (0.008) \\
\multirow{2}[0]{*}{ICLR} &     0.0283 &     0.0187 &     0.0197 &     0.0218 &     0.0197 &     0.0213 &     0.0210 \\
          &     (0.033) &     (0.033) &     (0.033) &     (0.033) &     (0.033) &     (0.033) &     (0.033) \\
\multirow{2}[0]{*}{ICML} &    -0.0068 &    -0.0122 &    -0.0119 &    -0.0132 &    -0.0117 &    -0.0112 &    -0.0123 \\
          &     (0.030) &     (0.030) &     (0.030) &     (0.030) &     (0.030) &     (0.030) &     (0.030) \\
\multirow{2}[0]{*}{N.Author} &     0.0361*** &     0.0271*** &     0.0338*** &     0.0145 &     0.0342*** &     0.0338*** &     0.0349*** \\
          &     (0.007) &     (0.008) &     (0.007) &     (0.009) &     (0.007) &     (0.007) &     (0.007) \\
\multirow{2}[0]{*}{Len.Abs.} &   9.49e-5** &  9.49e-5** &  9.44e-5** &  9.73e-5** &  9.43e-5** &  9.47e-5** &  9.42e-2** \\
          &  (3.57e-5) &  (3.6e-5) &  (3.58e-5) &  (3.65e-5) &  (3.6e-5) &  (3.59e-5) &  (3.58e-5) \\
\multirow{2}[0]{*}{Len.Title} &    -0.0011 &    -0.0011 &    -0.0011 &    -0.0012* &    -0.0011 &    -0.0011 &    -0.0011 \\
          &     (0.001) &     (0.001) &     (0.001) &     (0.001) &     (0.001) &     (0.001) &     (0.001) \\
\multirow{2}[0]{*}{Model} &    29.66*** &    29.55*** &    29.53*** &    29.84*** &    29.52*** &    29.54*** &    29.59*** \\
          &     (0.510) &     (0.509) &     (0.509) &     (0.504) &     (0.510) &     (0.510) &     (0.509) \\
\multirow{2}[0]{*}{HCTCD.Max} &     1.191 &       &       &       &       &       &  \\
          &     (0.752) &       &       &       &       &       &  \\
\multirow{2}[0]{*}{HCTCD.Min} &     0.1566 &       &       &       &       &       &  \\
          &     (0.993) &       &       &       &       &       &  \\
\multirow{2}[0]{*}{HCTCD.W.Sum} &       &     1.29*** &       &       &       &       &  \\
          &       &     (0.189) &       &       &       &       &  \\
\multirow{2}[0]{*}{HCTCD.wsum.d} &       &     0.9827*** &       &       &       &       &  \\
          &       &     (0.286) &       &       &       &       &  \\
\multirow{2}[0]{*}{HCTCD.Wave} &       &       &     2.66*** &       &       &       &  \\
          &       &       &     (0.384) &       &       &       &  \\
\multirow{2}[0]{*}{HCTCD.Wave.d} &       &       &     2.173*** &       &       &       &  \\
          &       &       &     (0.578) &       &       &       &  \\
\multirow{2}[0]{*}{HCTCD.Sum} &       &       &       &     0.4823*** &       &       &  \\
          &       &       &       &     (0.086) &       &       &  \\
\multirow{2}[0]{*}{HCTCD.Sum.d} &       &       &       &     0.3154* &       &       &  \\
          &       &       &       &     (0.138) &       &       &  \\
\multirow{2}[0]{*}{HCTCD.Ave.} &       &       &       &       &     2.753*** &       &  \\
          &       &       &       &       &     (0.430) &       &  \\
\multirow{2}[0]{*}{HCTCD.Ave.d} &       &       &       &       &     2.252*** &       &  \\
          &       &       &       &       &     (0.628) &       &  \\
\multirow{2}[0]{*}{HCTCD.1st} &       &       &       &       &       &     2.829*** &  \\
          &       &       &       &       &       &     (0.415) &  \\
\multirow{2}[0]{*}{HCTCD.1st.d} &       &       &       &       &       &     2.275*** &  \\
          &       &       &       &       &       &     (0.610) &  \\
\multirow{2}[0]{*}{HCTCD.Last} &       &       &       &       &       &       &     2.4335*** \\
          &       &       &       &       &       &       &     (0.366) \\
\multirow{2}[0]{*}{HCTCD.Last.d} &       &       &       &       &       &       &     2.0586*** \\
          &       &       &       &       &       &       &     (0.553) \\
\multirow{2}[1]{*}{Precision} &     1.7373*** &     1.7402*** &     1.7404*** &     1.7366*** &     1.7404*** &     1.7396*** &     1.7395*** \\
          &     (0.020) &     (0.020) &     (0.020) &     (0.020) &     (0.020) &     (0.020) &     (0.020) \\
    \midrule
    R-Squared & 0.6082 & 0.6091 & 0.6092 & 0.6081 & 0.609 & 0.6086 & 0.6088 \\
      Log-LL    &   1972.3 &   1978.6 &   1979.1 &   1970.2 & 1979.1 &   1977.6 &   1977.0 \\
      AIC              & -3923 & -3935 & -3936 & -3918 & -3936 & -3933 & -3932 \\
      BIC               & -3852 & -3865 & -3866 & -3848 & -3866 & -3863 & -3862 \\
    \bottomrule
    \end{tabular}%
            \begin{tablenotes}
      \item \textbf{Notes:} ***, **, and * indicate 0.001, 0.01, and 0.05 levels of significance, respectively.
    \end{tablenotes}
  \end{threeparttable}
  \label{tab:aggregation}}%
\end{table}%

Beta regression results are shown in Table ~\ref{tab:centralities}, where the precision parameter $\phi$ of beta regression is also estimated during the model fitting. They demonstrate significant differential impacts of centrality metrics on citation percentiles after controlling for covariates (venue, publication year, title/abstract length, and transformer-based content features). Closeness centrality exhibits the strongest positive association ($\hat{\gamma} = 1.44, p<0.001$), followed by summed degree centrality (
$\hat{\gamma} = 38.35, p<0.001$ and betweenness centrality ($\hat{\gamma} = 8.17, p<0.001$). Conversely, PageRank centrality displays paradoxical effects with a large coefficient ($\hat{\gamma} = 207.8, p<0.001$) but negligible practical significance due to multicollinearity with degree centrality ($r=0.961$). It is important to note that the magnitude of the fitted coefficients reflects both the association strength and the scale of the centrality metric distributions. Temporal differences of centrality ($X.d_i$) are statistically significant in all cases but consistently demonstrate weaker impact compared to cumulative metrics. Model fit substantially improves with centrality inclusion: The closeness-augmented model (Model 4) achieves $\Delta AIC=-37$ versus the baseline ($\Delta R^2=0.0032$), confirming centrality’s explanatory power beyond content and contextual features. Control variables align with expectations—citation percentiles decline over time ($\hat{\gamma}_1=-0.0374, p<0.001$) and scale with team size ($\hat{\gamma}_4=0.0378, p<0.001$). Precision parameters ($\phi \approx 1.73$) validate the beta distribution’s suitability for bounded citation data.

Table  ~\ref{tab:aggregation} further quantifies the efficacy of centrality aggregation methods, revealing that collective team connectivity drives citation advantage more than individual prominence. Summed closeness centrality across all authors ($\hat{\gamma}_{HCTCD.sum}=0.4823, p<0.001; R^2 = 0.6092$) outperforms both rank-based metrics (e.g., last-author centrality: $\hat{\gamma}_{HCTCD.last}=2.4335, p<0.001$) and extreme-value aggregations (max/min centrality: $p>0.05$). Weighted summation (exponentially discounting by author order) achieves the highest model fit ($\hat{\gamma}_{HCTCD.wsum}=1.29, p<0.001, AIC=-3935$). Notably, temporal derivatives of aggregated centrality (e.g., HCTCD.Sum.d) retain significance but contribute marginally to explanatory power ($\Delta R^2<0.001$). These findings robustly challenge conventional evaluation paradigms that emphasize lead authors or “star” researchers rather than holistic team integration within collaborative networks.


\subsection{The impact of incorporating high-centrality co-authors}
\begin{table}[htbp]
  \centering{\footnotesize
  \caption{Summary table of Beta regression on different centrality aggregation settings.}
  \begin{threeparttable}
    \begin{tabular}{p{2.0cm}p{1.7cm}p{1.6cm}p{1.6cm}p{1.5cm}p{1.5cm}p{1.5cm}p{1.5cm}}
    \toprule
    \textbf{Models} & \textbf{1} & \textbf{2} & \textbf{3} & \textbf{4} & \textbf{5} & \textbf{6} & \textbf{7} \\
    \midrule
    \multirow{2}[1]{*}{Const} &    -0.1555 & -0.163 &    -0.349*** &   -15.49*** &   -15.49*** &   -15.51*** &   -15.51*** \\
          &     (0.106) &     (0.106) &     (0.109) &     (0.278) &     (0.277) &     (0.277) &     (0.277) \\
    \multirow{2}[0]{*}{YearToNow} &    -0.0028 &    -0.0051 &    -0.0035 &    -0.0152 &    -0.0168* &    -0.0153 &    -0.0155 \\
          &     (0.010) &     (0.010) &     (0.010) &     (0.008) &     (0.009) &     (0.008) &     (0.008) \\
    \multirow{2}[0]{*}{ICLR} &     0.4077*** &     0.4038*** &     0.4120*** &     0.0154 & 0.0134 &     0.0195 &     0.0190 \\
          &     (0.040) &     (0.040) &     (0.040) &     (0.033) &     (0.033) &     (0.033) &     (0.033) \\
    \multirow{2}[0]{*}{ICML} &    -0.199*** &    -0.200*** &    -0.182*** &    -0.0200 &    -0.0211 &    -0.0125 &    -0.0124 \\
          &     (0.036) &   (0.036)  &   (0.036) &   (0.030) &     (0.030) &     (0.030) &     (0.030) \\
    \multirow{2}[0]{*}{Len.Abs.} &     0.0002*** &     0.0002*** &     0.0002*** &     0.0001** &     0.0001** &  9.36e-5** &  9.35e-5** \\
          &  (4.74e-5) &  (4.74e-5) & (4.72e-5) &  (3.56e-5) &  (3.54e-5) &  (3.58e-5) &  (3.58e-5) \\
    \multirow{2}[0]{*}{Len.Title} &    -0.006*** &    -0.006*** &    -0.006*** &    -0.0010 &    -0.0010 &    -0.0010 &    -0.0010 \\
          &     (0.001) &     (0.001) &     (0.001) &     (0.001) &     (0.001) &     (0.001) &     (0.001) \\
    \multirow{2}[0]{*}{Model} &       &       &       &    29.70*** &    29.69*** &    29.53*** &    29.53*** \\
          &       &       &       &     (0.510) &     (0.509) &     (0.510) &     (0.510) \\
    \multirow{2}[0]{*}{HCTCD.1st} &     8.77*** &     8.78*** &     7.71*** &     3.36*** &     3.37*** &     2.86*** &     2.82*** \\
          &     (0.479) &     (0.480) &     (0.503) &     (0.399) &     (0.399) &     (0.415) &     (0.428) \\
    \multirow{2}[0]{*}{HCTCD.1st.d} &     5.171*** &     4.911*** &     4.18*** &     2.675*** &     2.489*** &     2.123*** &     2.110*** \\
          &     (0.743) &     (0.749) &     (0.754) &     (0.605) &     (0.609) &     (0.614) &     (0.617) \\
    \multirow{2}[0]{*}{HCTCD.max.ind} &       &     0.1519** &     0.1278* &       &     0.1099* &     0.0976* &     0.0854 \\
          &       &     (0.053) &     (0.053) &       &     (0.044) &     (0.044) &     (0.052) \\
    \multirow{2}[0]{*}{N.Author} &       &       &     0.0629*** &       &       &     0.0327*** &     0.0327*** \\
          &       &       &     (0.009) &       &       &     (0.007) &     (0.008) \\
    \multirow{2}[0]{*}{HCTCD.1st.int} &       &       &       &       &       &       &     2.3621 \\
          &       &       &       &       &       &       &     (3.881) \\
    \multirow{2}[0]{*}{HCTCD.max.int} &       &       &       &       &       &       &    -1.0816 \\
          &       &       &       &       &       &       &     (2.006) \\
    \multirow{2}[1]{*}{Precision} &     1.083*** &     1.085*** &     1.097*** &     1.734*** &     1.736*** &     1.741*** &     1.741*** \\
          &     (0.019) &     (0.019) &     (0.019) &     (0.020) &     (0.020) &     (0.020) &     (0.020) \\
    \midrule
      Log-Likelihood   &   588.03 &   592.08 &   616.30 &   1967.4 &   1970.5 &   1980.1 &   1980.3 \\
      AIC    &   -1158. &   -1164. &   -1211. &   -3915. &   -3919. &   -3936. &   -3933. \\
      BIC  &   -1101. &   -1100. &   -1140. &   -3851. &   -3849. &   -3860. &   -3843. \\
    R2    & 0.2283 & 0.2301 & 0.241 & 0.6057 & 0.6065 & 0.6065 & 0.6094 \\
    \bottomrule
    \end{tabular}%
            \begin{tablenotes}
      \item \textbf{Notes:} ***, **, and * indicate 0.001, 0.01, and 0.05 levels of significance, respectively.
    \end{tablenotes}
  \end{threeparttable}
  \label{tab:high_coauthor}}%
\end{table}%
We perform further statistical analysis on the effect of including high-centrality co-authors given the first-author. We create a novel variable $X_{maxind}$, which get a value of $1$ if $X_{max}>(1+\alpha)X_{1st}$. We set $\alpha=0.5$ which means when a coauthor has a centrality value that is 50\% higher than the first author, the indicator equals to 1, otherwise equals to 0. In addition, we consider two interaction terms, the first is interaction between $X_{maxind}$ and $X_{1st}$, and the second is the interaction between $X_{maxind}$ and $X_{max}$. The corresponding equation is given by:
\begin{equation}
\begin{split}
u_i &= \gamma_0 + \gamma_1 \text{YearToNow}_i + \gamma_2 \text{ICLR}_i + \gamma_3 \text{ICML}_i + \gamma_4 N.Author + \gamma_5 \text{Len.Abs}_i \nonumber \\
&+ \gamma_6 \text{Len.Title}_i
 + \gamma_7 \text{Model}_i + \gamma_8 X_{1st,i} + \gamma_9 X.d._{1st,i} + \gamma_{10} X_{maxind,i} \\
 & + \gamma_{11} X_{maxind,i}\cdot X_{1st,i} + \gamma_{12}X_{maxind,i}\cdot X_{max,i} + \epsilon_i \nonumber
\end{split}
\end{equation}
We use the proposed centrality metric HCTCD as the centrality in the fitted model. The results of beta regression is given by Table~\ref{tab:high_coauthor}. We can see that after controlling the textual contents, the number of authors and the first author's centrality, incorporating a coauthor with a centrality that is higher than the first author by $50\%$ significantly associated with a increase of citation percentile ($0.08<\hat{\gamma}_{10}<0.16$). In addition, this effect is not significantly associated with the absolute centrality of the first author or the author with the maximum centrality in the author list, which is indicated by the insignificance of the interaction term coefficients $\gamma_{11}$ and $\gamma_{12}$. It is also noticeable that after controlling the first author and the maximum-centrality author, the number of authors is still significantly and positively associated with the citation percentile ($0.03<\hat{\gamma_4}<0.06$, $p<0.0001$).

\subsection{Likelihood ratio test of centrality impact}

To rigorously quantify the statistical significance of centrality metrics in explaining variation in citation percentiles, we conducted likelihood ratio tests (LRTs) comparing nested regression models. Specifically, we evaluated the full model, which incorporated both the time-aggregated centrality metric $X_{wave}$ and its temporal difference $X.d_{wave}$ against two reduced models: $\textit{Reduced}_1$, excluding $X.d_{wave}$, and $\textit{Reduced}_2$, excluding both $X_{wave}$ and 
$X.d_{wave}$. The results (Table~\ref{tab:lrt}) demonstrate that both the baseline centrality measure and its temporal dynamics significantly contribute to explanatory power. The test statistic for $\textit{Reduced}_1$ versus the full model was $LRT_{stat}=31.056$ which gives $p<0.001$, while $\textit{Reduced}_2$ versus the full model yielded $LRT_{stat}=339.637$, with $p<<0.001$. These outcomes robustly confirm that centrality metrics and their evolution over time are non-redundant predictors of citation disparities, reinforcing the critical role of sustained network positioning in scholarly impact.
\begin{table}[htbp]
  \centering{\small
  \caption{Likelihood ratio tests on the association of centrality metric and temporal difference of centrality metric.}
    \begin{tabular}{p{8em}cccc}
    \toprule
    Comparison & \multicolumn{1}{p{4.055em}}{$LL_{reduced}$} & \multicolumn{1}{p{4.055em}}{$LL_{full}$} & \multicolumn{1}{p{4.055em}}{$LRT_{Stat}$} & \multicolumn{1}{p{4.055em}}{p-value} \\
    \midrule
    $\text{Reduced}_1$ vs Full & 597.908  & 613.436  & 31.056  & 0.000  \\
    $\text{Reduced}_2$ vs Full & 443.617  & 613.436  & 339.637  & 0.000  \\
    \bottomrule
    \end{tabular}%
  \label{tab:lrt}}%
\end{table}%

\subsection{Predictive analysis}
We apply a combination of predictors including the centrality metrics, the year, venue and the output of pretrained transformer-based predictive models, and predict the percentile of the citations among all the papers published in the same year. We train models including linear regression, feedforward neural networks, random forest and XGBoost. We split the original datasets into training-validation set and test set by 0.9:0.1. We use 5-fold cross-validation within the training-validation set for model selection and hyper-parameter tuning, and the test set is used to evaluate the generalization performance of the trained model. We tune the major hyper-parameters for each model, the final hyper-parameters setting is selected as the setting gives the optimal cross-validation loss. For random forest, the number of estimators equals to 290, maximum depth equals to 15, and min samples leaf is 3. For XGBoost, the number of estimators equals to 178, maximum depth equals to 11, the learning rate is 0.03426 and colsample bytree is 0.7. For feedforward neural networks, we use three hidden layers with hidden size equals to 32, while the learning rate is 0.005 and the batch size is 64.

The results of different models on the test set are given in Table~\ref{tab:predictive}. It demonstrates that centrality metrics enhance citation percentile forecasting across machine learning models. Incorporating centrality reduces MSE by 2.4–4.8\% (e.g., XGBoost: MSE = 0.03502 vs. 0.03645 without centrality) and boosts correlation by 1.2–1.7\%. Random Forest achieves optimal correlation (Corr = 0.717) with centrality features, while Linear Regression shows the greatest relative improvement ($\Delta$Corr = 0.0083). These gains, though modest, confirm centrality’s value as a complementary predictor beyond content and venue features. The persistent advantage across heterogeneous algorithms (parametric vs. tree-based) underscores the robustness of network-derived signals in impact prediction.
\begin{table}[htbp]
  \centering{\small
  \caption{Results of Predictive Performance of Machine Learning Models.}
    \begin{tabular}{ccccccc}
    \toprule
          & \multicolumn{3}{c}{w/ Centralities} & \multicolumn{3}{c}{w/o Centralities} \\
          & MSE   & MAE   & Corr. & MSE   & MAE   & Corr. \\
    \midrule
    LR    & 0.03529 & 0.1484 & 0.7142 & 0.03616 & 0.1498 & 0.7059 \\
    Random Forest & 0.035213 & 0.1452 & 0.717 & 0.03698 & 0.1497 & 0.7004 \\
    XGBoost & 0.03502 & 0.14711 & 0.71647 & 0.03645 & 0.1505 & 0.7025 \\
    Neural Networks & 0.0357 & 0.1498 & 0.7101 & 0.03678 & 0.1514 & 0.7025 \\
    \bottomrule
    \end{tabular}%
  \label{tab:predictive}}%
\end{table}%

\section{Discussion}

\subsection{Compare with related studies}

We further compare our results with existing studies. As we know, almost all existing studies fail to control the textual contents with pre-trained AI models, and they did not evaluate the impact of adding high-centrality authors given the first author (after controlling the contents). Beyond that, \citet{yan2009applying} demonstrated significant correlations between various centrality measures (closeness, betweenness, degree, and PageRank) and citation counts in library and information science journals, yet their explanatory power remained limited, likely due to the domain difference, data preprocessing techniques, and the omission of temporal dynamics and team-level aggregation. Our study advances this line of inquiry by introducing a temporal-decay adjusted centrality metric (HCTCD) and demonstrating that long-term centrality (e.g., 16-year windows) exerts a stronger influence than short-term measures, which is a nuance not captured in earlier work. 
Moreover, \citet{guan2017impact} integrated knowledge networks with collaboration networks to explain citations, yet their models did not fully account for the bounded nature of citation data or the aggregation strategies across author lists. Our use of beta regression to model citation percentiles addresses this methodological gap, providing a more robust estimation framework than ordinary least squares regression commonly used in earlier studies. Additionally, while \citet{sarigol2014predicting} reported a positive correlation between author centrality and citation success in computer science, they did not explore the differential impacts of centrality aggregation methods at the team level. Our finding that weighted summation of centrality values across all authors outperforms rank-based approaches (e.g., first- or last-author centrality) challenges conventional evaluation paradigms and aligns with emerging emphasis on collective rather than individual metrics in impact assessment.

\subsection{Theoretical insight}
While our study empirically establishes a significant relationship between author centrality and citation outcomes, it is essential to situate these findings within broader theoretical frameworks to elucidate the underlying mechanisms. From a social capital theory perspective, authors embedded in central network positions accumulate relational resources that enhance the visibility, credibility, and diffusion of their work \citep{holes1992social, nahapiet1998social}. High-centrality authors often possess greater access to diverse knowledge sources and influential audiences, which accelerates the dissemination and recognition of their publications. Complementary to this, knowledge diffusion theory posits that network centrality reduces the path length of information flow, enabling research to reach wider audiences more efficiently \citep{valente1996network, rogers2014diffusion}. Moreover, signaling theory suggests that authorship by well-connected scholars may act as a quality cue, leading to higher citation rates due to perceived credibility \citep{connelly2011signaling, siler2022cumulative}. Consequently, controlling for topic and content is indispensable when isolating the true influence of centrality from variations in intrinsic quality. Integrating these perspectives, we contend that centrality operates not merely as a structural feature but as a proxy for unobserved social and cognitive processes that drive citation disparities. This theoretical grounding underscores the importance of moving beyond purely bibliometric indicators toward a more nuanced understanding of how network embeddedness shapes scholarly impact.

\subsection{Implication of research management} \label{Sec:5.3}

Our findings offer actionable insights for reforming research evaluation practices across multiple domains: 

\textbf{Funding Allocation}:
Funding agencies should place greater emphasis on supporting research teams that demonstrate strong and diverse collaborative networks, rather than those overly reliant on a single high-profile researcher. To operationalize this, agencies could incorporate team-based network metrics (such as our proposed HCTCD) into evaluation criteria. This approach helps counter cumulative advantage biases and promotes a more equitable distribution of research resources.

\textbf{Academic Promotion}:
Institutions are suggested to redesign promotion frameworks to weight collective network metrics alongside traditional output-based indicators. For instance, evaluating collaborative network measures and intellectual contribution metrics as two distinct dimensions, rather than relying solely on first- or corresponding-author status, which can provide a more holistic assessment of an individual’s collaborative impact and reduce over-reliance on symbolic authorship.

\textbf{Journal Review Process}:
Journals and conferences could implement ``network-aware'' review protocols, leveraging centrality analytics to identify potential honorary authorships. Tools that flag papers with high centrality disparities among co-authors may prompt editors to request explicit contribution statements, thereby enhancing accountability and reducing gratuitous authorship.

These interventions, grounded in empirical network analysis, hold the potential to contribute to a more efficient, transparent, and equitable scientific ecosystem.

\section{Conclusion}

This research demonstrates that authors’ enduring positions within scientific collaboration networks significantly drive systemic disparities in citation outcomes after controlling the research topic or content. Through longitudinal analysis of 17,942 machine learning publications, we establish that long-term centrality (e.g., 16-year closeness centrality) exerts stronger influence on citations than short-term collaboration patterns. Methodologically, we advance citation analysis by developing the HCTCD centrality metric, which incorporates temporal decay and collaboration intensity. Meanwhlie, we employ Beta regression to model bounded citation percentiles, which is a statistically rigorous approach that outperforms traditional OLS for $[0,1]$-distributed data. Critically, our models control for textual content via transformer-based embeddings, confirming that centrality effects persist independently of research substance. We further reveal that given the paper topic and the first author, involving co-authors with high centralities (50\% higher relative to the first author) significantly elevates a paper’s expected citation percentile, underscoring how honorary authorship structurally advantages well-connected teams.

As is discussed in Section~\ref{Sec:5.3}, these findings suggest stakeholders to reform research evaluation practices. Funding agencies should prioritize teams exhibiting diversified network integration over those reliant on isolated ``star'' researchers, thereby mitigating cumulative advantage biases in resource allocation. Journals and conferences could implement ``network-aware'' review protocols, auditing citation networks to discount honorary authorship influence. Academic institutions are suggested to redesign promotion frameworks by weighting collaboration/centrality metrics alongside contribution metrics. While our focus on machine learning invites future validation in interdisciplinary contexts, the consistent dominance of team centrality aggregation over individual-author metrics fundamentally challenges entrenched evaluation paradigms.

We acknowledge limitations regarding domain generalizability and causal identification. However, the robust predictive gains from centrality features in machine learning models validate their utility for academic governance. Moving forward, translating these insights into equity-centered interventions (e.g., restricting high-impact researchers from being listed as authors without substantive contributions, and ensure that all authors participate meaningfully in the research and take responsibility for the quality of the paper) promises to foster a more efficient and equitable scientific ecosystem. By bridging network theory, scientometrics, and management science, this work reorients citation equity from abstract critique to actionable reform.

\bibliography{Reference}

@book{aldrich1984linear,
  title={Linear probability, logit, and probit models},
  author={Aldrich, John H and Nelson, Forrest D},
  number={45},
  year={1984},
  publisher={Sage}
}

@inproceedings{yu2012citation,
  title={Citation prediction in heterogeneous bibliographic networks},
  author={Yu, Xiao and Gu, Quanquan and Zhou, Mianwei and Han, Jiawei},
  booktitle={Proceedings of the 2012 SIAM International Conference on Data Mining},
  pages={1119--1130},
  year={2012},
  organization={SIAM}
}

@inproceedings{geng2022modeling,
  title={Modeling dynamic heterogeneous graph and node importance for future citation prediction},
  author={Geng, Hao and Wang, Deqing and Zhuang, Fuzhen and Ming, Xuehua and Du, Chenguang and Jiang, Ting and Guo, Haolong and Liu, Rui},
  booktitle={Proceedings of the 31st ACM International Conference on Information \& Knowledge Management},
  pages={572--581},
  year={2022}
}

@inproceedings{yang2023revisiting,
  title={Revisiting citation prediction with cluster-aware text-enhanced heterogeneous graph neural networks},
  author={Yang, Carl and Han, Jiawei},
  booktitle={2023 IEEE 39th International Conference on Data Engineering (ICDE)},
  pages={682--695},
  year={2023},
  organization={IEEE}
}

@inproceedings{bhat2015citation,
  title={Citation prediction using diverse features},
  author={Bhat, Harish S and Huang, Li-Hsuan and Rodriguez, Sebastian and Dale, Rick and Heit, Evan},
  booktitle={2015 IEEE International Conference on Data Mining Workshop (ICDMW)},
  pages={589--596},
  year={2015},
  organization={IEEE}
}

@article{iqbal2021decade,
  title={A decade of in-text citation analysis based on natural language processing and machine learning techniques: an overview of empirical studies},
  author={Iqbal, Sehrish and Hassan, Saeed-Ul and Aljohani, Naif Radi and Alelyani, Salem and Nawaz, Raheel and Bornmann, Lutz},
  journal={Scientometrics},
  volume={126},
  number={8},
  pages={6551--6599},
  year={2021},
  publisher={Springer}
}

@article{petersen2015quantifying,
  title={Quantifying the impact of weak, strong, and super ties in scientific careers},
  author={Petersen, Alexander Michael},
  journal={Proceedings of the National Academy of Sciences},
  volume={112},
  number={34},
  pages={E4671--E4680},
  year={2015},
  publisher={National Academy of Sciences}
}

@article{nielsen2021global,
  title={Global citation inequality is on the rise},
  author={Nielsen, Mathias Wullum and Andersen, Jens Peter},
  journal={Proceedings of the National Academy of Sciences},
  volume={118},
  number={7},
  pages={e2012208118},
  year={2021},
  publisher={National Academy of Sciences}
}

@article{lerman2022gendered,
  title={Gendered citation patterns among the scientific elite},
  author={Lerman, Kristina and Yu, Yulin and Morstatter, Fred and Pujara, Jay},
  journal={Proceedings of the National Academy of Sciences},
  volume={119},
  number={40},
  pages={e2206070119},
  year={2022},
  publisher={National Academy of Sciences}
}

@article{liu2023non,
  title={Non-White scientists appear on fewer editorial boards, spend more time under review, and receive fewer citations},
  author={Liu, Fengyuan and Rahwan, Talal and AlShebli, Bedoor},
  journal={Proceedings of the National Academy of Sciences},
  volume={120},
  number={13},
  pages={e2215324120},
  year={2023},
  publisher={National Academy of Sciences}
}

@article{catalini2015incidence,
  title={The incidence and role of negative citations in science},
  author={Catalini, Christian and Lacetera, Nicola and Oettl, Alexander},
  journal={Proceedings of the National Academy of Sciences},
  volume={112},
  number={45},
  pages={13823--13826},
  year={2015},
  publisher={National Academy of Sciences}
}

@article{albert2002statistical,
  title={Statistical mechanics of complex networks},
  author={Albert, R{\'e}ka and Barab{\'a}si, Albert-L{\'a}szl{\'o}},
  journal={Reviews of Modern Physics},
  volume={74},
  number={1},
  pages={47},
  year={2002},
  publisher={APS}
}

@article{newman2003structure,
  title={The structure and function of complex networks},
  author={Newman, Mark EJ},
  journal={SIAM Review},
  volume={45},
  number={2},
  pages={167--256},
  year={2003},
  publisher={SIAM}
}

@article{varga2019shorter,
  title={Shorter distances between papers over time are due to more cross-field references and increased citation rate to higher-impact papers},
  author={Varga, Attila},
  journal={Proceedings of the National Academy of Sciences},
  volume={116},
  number={44},
  pages={22094--22099},
  year={2019},
  publisher={National Academy of Sciences}
}

@article{radicchi2008universality,
  title={Universality of citation distributions: Toward an objective measure of scientific impact},
  author={Radicchi, Filippo and Fortunato, Santo and Castellano, Claudio},
  journal={Proceedings of the National Academy of Sciences},
  volume={105},
  number={45},
  pages={17268--17272},
  year={2008},
  publisher={National Academy of Sciences}
}

@article{wislar2011honorary,
  title={Honorary and ghost authorship in high impact biomedical journals: a cross sectional survey},
  author={Wislar, Joseph S and Flanagin, Annette and Fontanarosa, Phil B and DeAngelis, Catherine D},
  journal={Bmj},
  volume={343},
  year={2011},
  publisher={British Medical Journal Publishing Group}
}

@misc{anstey2014authorship,
  title={Authorship issues: grizzles, guests and ghosts},
  author={Anstey, Alex},
  journal={British Journal of Dermatology},
  volume={170},
  number={6},
  pages={1209--1210},
  year={2014},
  publisher={Blackwell Publishing Ltd Oxford, UK}
}

@article{morreim2023guest,
  title={Guest authorship as research misconduct: definitions and possible solutions},
  author={Morreim, EH and Winer, Jeffrey C},
  journal={BMJ Evidence-Based Medicine},
  volume={28},
  number={1},
  pages={1--4},
  year={2023},
  publisher={Royal Society of Medicine}
}

@article{flanagin1998prevalence,
  title={Prevalence of articles with honorary authors and ghost authors in peer-reviewed medical journals},
  author={Flanagin, Annette and Carey, Lisa A and Fontanarosa, Phil B and Phillips, Stephanie G and Pace, Brian P and Lundberg, George D and Rennie, Drummond},
  journal={Jama},
  volume={280},
  number={3},
  pages={222--224},
  year={1998},
  publisher={American Medical Association}
}

@article{liu2018review,
  title={A review and comparison of Bayesian and likelihood-based inferences in beta regression and zero-or-one-inflated beta regression},
  author={Liu, Fang and Eugenio, Evercita C},
  journal={Statistical Methods in Medical Research},
  volume={27},
  number={4},
  pages={1024--1044},
  year={2018},
  publisher={SAGE Publications Sage UK: London, England}
}

@article{ferrari2004beta,
  title={Beta regression for modelling rates and proportions},
  author={Ferrari, Silvia and Cribari-Neto, Francisco},
  journal={Journal of Applied Statistics},
  volume={31},
  number={7},
  pages={799--815},
  year={2004},
  publisher={Taylor \& Francis}
}

@article{camarinha2005collaborative,
  title={Collaborative networks: a new scientific discipline},
  author={Camarinha-Matos, Luis M and Afsarmanesh, Hamideh},
  journal={Journal of Intelligent Manufacturing},
  volume={16},
  pages={439--452},
  year={2005},
  publisher={Springer}
}

@inproceedings{tang2016aminer,
  title={AMiner: Toward understanding big scholar data},
  author={Tang, Jie},
  booktitle={Proceedings of the Ninth ACM International Conference on Web Search and Data Mining (WSDM)},
  pages={467--467},
  year={2016}
}

@article{wan2019aminer,
  title={Aminer: Search and mining of academic social networks},
  author={Wan, Huaiyu and Zhang, Yutao and Zhang, Jing and Tang, Jie},
  journal={Data Intelligence},
  volume={1},
  number={1},
  pages={58--76},
  year={2019},
  publisher={MIT Press One Rogers Street, Cambridge, MA 02142-1209, USA journals-info~…}
}

@article{sebah2002introduction,
  title={Introduction to the gamma function},
  author={Sebah, Pascal and Gourdon, Xavier},
  journal={American Journal of Scientific Research},
  volume={2},
  year={2002}
}

@article{kwon2022rise,
  title={The rise of citational justice: how scholars are making references fairer.},
  author={Kwon, Diana},
  journal={Nature},
  volume={603},
  number={7902},
  pages={568--571},
  year={2022}
}

@article{deng2015papers,
  title={Papers with shorter titles get more citations},
  author={Deng, Boer},
  journal={Nature},
  volume={2},
  number={8},
  pages={150266},
  year={2015}
}

@misc{adams2014bibliometrics,
  title={Bibliometrics: The citation game},
  author={Adams, Jonathan},
  journal={Nature},
  year={2014},
  publisher={Nature Publishing Group UK London}
}

@article{teich2022citation,
  title={Citation inequity and gendered citation practices in contemporary physics},
  author={Teich, Erin G and Kim, Jason Z and Lynn, Christopher W and Simon, Samantha C and Klishin, Andrei A and Szymula, Karol P and Srivastava, Pragya and Bassett, Lee C and Zurn, Perry and Dworkin, Jordan D and others},
  journal={Nature Physics},
  volume={18},
  number={10},
  pages={1161--1170},
  year={2022},
  publisher={Nature Publishing Group UK London}
}

@inproceedings{zhang2017degree,
  title={Degree centrality, betweenness centrality, and closeness centrality in social network},
  author={Zhang, Junlong and Luo, Yu},
  booktitle={2017 2nd International Conference on Modelling, Simulation and applied Mathematics (MSAM2017)},
  pages={300--303},
  year={2017},
  organization={Atlantis press}
}

@article{freeman2002centrality,
  title={Centrality in social networks: Conceptual clarification},
  author={Freeman, Linton C and others},
  journal={Social Network: Critical Concepts in Sociology. Londres: Routledge},
  volume={1},
  number={3},
  pages={238--263},
  year={2002}
}

@techreport{page1999pagerank,
  title={The PageRank citation ranking: Bringing order to the web.},
  author={Page, Lawrence and Brin, Sergey and Motwani, Rajeev and Winograd, Terry},
  year={1999},
  institution={Stanford infolab}
}

@article{sarigol2014predicting,
  title={Predicting scientific success based on coauthorship networks},
  author={Sarig{\"o}l, Emre and Pfitzner, Ren{\'e} and Scholtes, Ingo and Garas, Antonios and Schweitzer, Frank},
  journal={EPJ Data Science},
  volume={3},
  pages={1--16},
  year={2014},
  publisher={Springer}
}

@article{xia2023review,
  title={A review of scientific impact prediction: tasks, features and methods},
  author={Xia, Wanjun and Li, Tianrui and Li, Chongshou},
  journal={Scientometrics},
  volume={128},
  number={1},
  pages={543--585},
  year={2023},
  publisher={Springer}
}

@article{guan2017impact,
  title={The impact of collaboration and knowledge networks on citations},
  author={Guan, Jiancheng and Yan, Yan and Zhang, Jing Jing},
  journal={Journal of Informetrics},
  volume={11},
  number={2},
  pages={407--422},
  year={2017},
  publisher={Elsevier}
}

@article{yan2009applying,
  title={Applying centrality measures to impact analysis: A coauthorship network analysis},
  author={Yan, Erjia and Ding, Ying},
  journal={Journal of the American Society for Information Science and Technology},
  volume={60},
  number={10},
  pages={2107--2118},
  year={2009},
  publisher={Wiley Online Library}
}

@article{biscaro2014co,
  title={Co-authorship and bibliographic coupling network effects on citations},
  author={Biscaro, Claudio and Giupponi, Carlo},
  journal={PloS one},
  volume={9},
  number={6},
  pages={e99502},
  year={2014},
  publisher={Public Library of Science San Francisco, USA}
}

@article{nahapiet1998social,
  title={Social capital, intellectual capital, and the organizational advantage},
  author={Nahapiet, Janine and Ghoshal, Sumantra},
  journal={Academy of Management Review},
  volume={23},
  number={2},
  pages={242--266},
  year={1998},
  publisher={Academy of Management Briarcliff Manor, NY 10510}
}

@book{holes1992social,
  title={The social structure of competition},
  author={Holes, Structural},
  year={1992},
  publisher={Cambridge, MA: Harvard University Press}
}

@misc{valente1996network,
  title={Network models of the diffusion of innovations},
  author={Valente, Thomas W},
  year={1996},
  publisher={Springer}
}

@incollection{rogers2014diffusion,
  title={Diffusion of innovations},
  author={Rogers, Everett M and Singhal, Arvind and Quinlan, Margaret M},
  booktitle={An integrated approach to communication theory and research},
  pages={432--448},
  year={2014},
  publisher={Routledge}
}

@article{connelly2011signaling,
  title={Signaling theory: A review and assessment},
  author={Connelly, Brian L and Certo, S Trevis and Ireland, R Duane and Reutzel, Christopher R},
  journal={Journal of Management},
  volume={37},
  number={1},
  pages={39--67},
  year={2011},
  publisher={Sage Publications Sage CA: Los Angeles, CA}
}

@article{siler2022cumulative,
  title={Cumulative advantage and citation performance of repeat authors in scholarly journals},
  author={Siler, Kyle and Vincent-Lamarre, Philippe and Sugimoto, Cassidy R and Larivi{\`e}re, Vincent},
  journal={Plos One},
  volume={17},
  number={4},
  pages={e0265831},
  year={2022},
  publisher={Public Library of Science San Francisco, CA USA}
}

@article{white1994betweenness,
  title={Betweenness centrality measures for directed graphs},
  author={White, Douglas R and Borgatti, Stephen P},
  journal={Social Networks},
  volume={16},
  number={4},
  pages={335--346},
  year={1994},
  publisher={Elsevier}
}

@article{brandes2001faster,
  title={A faster algorithm for betweenness centrality},
  author={Brandes, Ulrik},
  journal={Journal of Mathematical Sociology},
  volume={25},
  number={2},
  pages={163--177},
  year={2001},
  publisher={Taylor \& Francis}
}

@article{senanayake2015pagerank,
  title={The pagerank-index: Going beyond citation counts in quantifying scientific impact of researchers},
  author={Senanayake, Upul and Piraveenan, Mahendra and Zomaya, Albert},
  journal={PloS One},
  volume={10},
  number={8},
  pages={e0134794},
  year={2015},
  publisher={Public Library of Science San Francisco, CA USA}
}

@article{zhang2022analysing,
  title={Analysing academic paper ranking algorithms using test data and benchmarks: an investigation},
  author={Zhang, Yu and Wang, Min and Saberi, Morteza and Chang, Elizabeth},
  journal={Scientometrics},
  volume={127},
  number={7},
  pages={4045--4074},
  year={2022},
  publisher={Springer}
}

@article{kane2008casting,
  title={Casting the net: A multimodal network perspective on user-system interactions},
  author={Kane, Gerald C and Alavi, Maryam},
  journal={Information Systems Research},
  volume={19},
  number={3},
  pages={253--272},
  year={2008},
  publisher={INFORMS}
}

@article{sasidharan2012effects,
  title={The effects of social network structure on enterprise systems success: A longitudinal multilevel analysis},
  author={Sasidharan, Sharath and Santhanam, Radhika and Brass, Daniel J and Sambamurthy, Vallabh},
  journal={Information Systems Research},
  volume={23},
  number={3-part-1},
  pages={658--678},
  year={2012},
  publisher={INFORMS}
}

@article{nezami2025network,
  title={Network centrality and firm performance: A meta-analysis},
  author={Nezami, Mehdi and Chisam, Natalie and Palmatier, Robert W},
  journal={Journal of the Academy of Marketing Science},
  volume={53},
  number={1},
  pages={79--104},
  year={2025},
  publisher={Springer}
}

@article{he2022executive,
  title={Executive network centrality and corporate reporting},
  author={He, Jing},
  journal={Management Science},
  volume={68},
  number={2},
  pages={1512--1536},
  year={2022},
  publisher={INFORMS}
}

@article{ahuja2003individual,
  title={Individual centrality and performance in virtual R\&D groups: An empirical study},
  author={Ahuja, Manju K and Galletta, Dennis F and Carley, Kathleen M},
  journal={Management science},
  volume={49},
  number={1},
  pages={21--38},
  year={2003},
  publisher={INFORMS}
}

@article{freeman1978centrality,
  title={Centrality in social networks conceptual clarification},
  author={Freeman, Linton C},
  journal={Social Networks},
  volume={1},
  number={3},
  pages={215--239},
  year={1978},
  publisher={North-Holland}
}

@article{stephenson1989rethinking,
  title={Rethinking centrality: Methods and examples},
  author={Stephenson, Karen and Zelen, Marvin},
  journal={Social Networks},
  volume={11},
  number={1},
  pages={1--37},
  year={1989},
  publisher={Elsevier}
}

\end{document}